\documentclass[12pt]{article}
\usepackage{latexsym}
\usepackage{amssymb}
\usepackage{amsmath}
\usepackage{graphicx}
\usepackage{enumitem}
\usepackage{bbold}
\usepackage{slashed}
\usepackage{hyperref}

\newcommand\mathC{\mkern1mu\raise2.2pt\hbox{$\scriptscriptstyle|$}
        {\mkern-7mu\rm C}}              

\def\be{\begin{equation}}
\def\ee{\end{equation}}
\def\bear{\begin{eqnarray}}
\def\eear{\end{eqnarray}}
\def\nn{\nonumber}

\newcommand\bra[1]{{\langle {#1}|}}
\newcommand\ket[1]{{|{#1}\rangle}}

\def\b{\beta}

\def\t{\tau}

\def\s{\sigma}
\def\th{\theta}

\def\dd{\mbox{d}}

\def\bra{\langle}
\def\ket{\rangle}

\def\b{\beta}

\def\F{\Phi}

\def\l{\lambda}

\def\m{\mu}
\def\n{\nu}

\def\t{\tau}
\def\th{\theta}

\newcommand{\sm}[1]{\mbox{\scriptsize #1}}
\newcommand{\tn}[1]{\mbox{\tiny #1}}
\renewcommand{\@}[1]{\sqrt{#1}}
\renewcommand{\le}[1]{\label{#1}\end{eqnarray}}
\newcommand{\bea}{\begin{eqnarray}}
\newcommand{\eea}{\end{eqnarray}}

\newcommand{\eq}[1]{(\ref{#1})}
\def\nn{\nonumber\\}

\def\ffract#1#2{\raise .35 em\hbox{$\scriptstyle#1$}\kern-.25em/
\kern-.2em\lower .22 em \hbox{$\scriptstyle#2$}}

\begin{document}

\pagestyle{empty}

\centerline{{\Large \bf Towards a Theory of Emergence for the Physical Sciences}}
\vskip 1cm

\begin{center}
{\large Sebastian De Haro}\\
\vskip 1truecm
{\it Trinity College, Cambridge, CB2 1TQ, United Kingdom}\\
{\it Department of History and Philosophy of Science, University of Cambridge\\
Vossius Center for History of Humanities and Sciences, University of Amsterdam}\\

\vskip .7truecm
{\tt sd696@cam.ac.uk}
\vskip 1truecm
\today
\end{center}

\vskip 7truecm

\begin{center}
\textbf{\large \bf Abstract}
\end{center}

I begin to develop a framework for emergence in the physical sciences. Namely, I propose to explicate ontological emergence in terms of the notion of `novel reference', and of an account of interpretation as a map from theory to world. I then construe ontological emergence as the ``failure of the interpretation to mesh'' with an appropriate linkage map between theories. Ontological emergence can obtain between theories that have the same extension but different intensions, and between theories that have both different extensions and intensions. 

I illustrate the framework in three examples: the emergence of spontaneous magnetisation in a ferromagnet, the emergence of masslessness, and the emergence of space, in specific models of physics.

The account explains {\it why} ontological emergence is independent of reduction: namely, because emergence is primarily concerned with adequate {\it interpretation}, while the sense of reduction that is relevant here is concerned with inter-theoretic relations between {\it uninterpreted} theories.

\newpage
\pagestyle{plain}

\tableofcontents

\newpage

\section{Introduction}\label{intro}

The aim of this paper is to introduce and illustrate a criterion for ontological emergence. The framework is formal, where by `formal' I just mean `admitting of the basic notions of sets and maps'. The framework will then be illustrated by three examples: the emergence of spontaneous magnetization in a ferromagnet, the emergence of masslessness in classical relativistic mechanics, and the emergence of space in a random matrix model.

The paper can be seen as a straightforward explication of the phrase `ontological emergence'. Although my explication is related to other construals of this notion, in particular Humphreys (2016:~pp.~56-93), I believe that my construal contains a number of novel aspects, and that the way I formalise it---as a single non-meshing condition between two maps---is a helpful tool for analysing cases of emergence, and for further conceptual analysis. 
The non-meshing condition can be understood as a difference in the intensions of two theories related by linkage (and sometimes also the extensions differ).

My immediate aim here is not strongly metaphysical, in the sense of requiring a commitment to a specific metaphysics of the world, and explicating emergence in those terms. Rather, my aim is to clarify what we mean by the phrase `ontological emergence' (often contrasted with `epistemic emergence') {\it in general}: and to give a criterion that is as straightforward as possible---a sufficient condition---for when it occurs. Thus I aim to give a minimal account of the meaning of `ontological emergence', independently of whether we are e.g.~Humeans or Aristotelians about causation---further metaphysical details then just adding to the basic picture that I will present here.

Thus I will here construe `ontology' in the straightforward sense of `the ontology of a scientific theory', i.e.~the domain of application that a theory describes, under a given interpretation. This domain of application is a part of the empirical world. Thus ontology is here not understood as a piece of language, but as a part of the world (more about this in Section \ref{fdev}). 

I will also follow the new wave of emergentism in the physical sciences, which goes back to Anderson's (1972) `More is Different'. The new emergentists have focussed on the emergence of entities and of properties\footnote{See for example Laughlin and Pines (2000:~p.~28), Anderson (1989:~p.~586). } rather than on for example causal powers or causal properties, which are notions that require further metaphysical explication. While questions of causation are important in e.g.~philosophy of mind, I follow the new emergentists in thinking that a minimal account of emergence can avoid them (see e.g.~Bedau (1997:~pp.~376-377), Hendry (2010:~pp.~184-185)). We will see that the first steps taken by the present framework already present a number of questions and themes worth clarifying in their own right. 

The writings of the new emergentists in physics are unfortunately not precise enough for us to extract from them a doctrine about emergence. The debate between emergentists and reductionists seems to have been fuelled by the alleged incompatibility between reduction and emergence.\footnote{The main opponent of Anderson's emergentist doctrine was Steven Weinberg (Bedau and Humphreys (2008:~pp.~345-357)). For a philosophical discussion of the debates, see Mainwood (2006: Section 2).} Thus part of my aim is to offer a framework for emergence that allows for the coexistence of emergence and (at least one widespread type of) reduction, the possibility of which has been cogently defended for example by Butterfield (2011, 2011a). 

An influential explication of emergence is by the British emergentist C.~D.~Broad (1925, p.~61): 

\begin{quote}\small Put in abstract terms the emergent theory asserts that there are certain wholes, composed (say) of constituents, $A$, $B$ and $C$ in a relation $R$ to each other; that all wholes composed of constituents of the same kind as $A$, $B$ and $C$ in relations of the same kind as $R$ have certain characteristic properties; that $A$, $B$ and $C$ are capable of occurring in other kinds of complex where the relation is not of the same kind as $R$; and that the characteristic properties of the whole $R(A;B; C)$ cannot, even in theory, be deduced from the most complete knowledge of the properties of $A$, $B$ and $C$ in isolation or in other wholes which are not of the form $R(A;B; C)$.
\end{quote}

While Broad's construal of `emergence' has been influential, I also submit that his description, and others that define ontological emergence as the `lack of deducibility', are too strong:\footnote{For a critical review of the old notion of `emergence as a failure of reduction' (where reduction is standardly defined as deduction, using ``bridge-laws'', {\it \`a la} Nagel), see Bedau and Humphreys (2008:~pp.~10-11), Humphreys (2016:~p.~37). For why emergence is compatible with reduction, see Section \ref{nornc}.} Broad (1925:~p.~59) himself acknowledged that his theory of emergence fell short of finding interesting empirical illustrations: `I cannot give a conclusive example of it, since it is a matter of controversy whether it actually applies to anything'. 
But we do not need to define emergence as lack of deducibility: what we need, I will argue, is to make a distinction between the {\it formalism} of a theory (which is the part of the theory that best allows us to use notions such as deduction) and the theory's {\it interpretation}, which is about the world, and need not be subject to such logical relations. This will allow a more general definition of emergence as novelty.

Another difficulty with the notion of emergence is that it is used broadly, and has various connotations. Guay and Sartenaer (2016) have recently carried out an interesting exercise in distinguishing three directions in the emergence landscape, through the contrasts: epistemological vs.~ontological, weak vs.~strong (i.e.~emergence `in practice' vs.~`in principle'), and synchronic vs.~diachronic emergence. In this paper, I will concentrate on ontological, synchronic emergence. 

Let me distinguish two main meanings of the word `formal' that I will use. The official meaning of `formal', as I announced at the beginning of this Introduction, is as in `mathematics applied to philosophy': more specifically, in the sense of applying the notions of sets and maps to physical theories to articulate how they denote items in the world and theory-world relations. Thus my explication of `emergence' is {\it formal} in that it applies to theories that are so formalised. What is `formal' in this main meaning can still be interpreted, i.e.~`formal' here does {\it not} contrast with interpretation and ontology. The second meaning of `formal' {\it does} contrast with `interpretation': for it denotes the formalism, or the mathematical formulation, of physical theories, stripped of their physical interpretations. Since my main meaning of `formal' is the former, I will use the phrases `formalism of a theory' or `formal, i.e.~not interpretative' when referring to the latter.

The paper proceeds as follows. Section \ref{theoryem} lays out the framework for emergence: including 
the conception of epistemic and ontological emergence. 
Section \ref{fdev} discusses three further questions that the framework prompts and compares with the literature. Sections \ref{imct}, \ref{mmpp}, and \ref{casemm} illustrate the framework in three case studies. Section \ref{conclusion} concludes.

\section{Towards a Theory of Emergence}\label{theoryem}

This Section develops the main framework of the paper. In Section \ref{ern}, I give the necessary background about emergence, reduction, approximation, and reference: which I will use, in Section \ref{onep}, to explicate ontological vs.~epistemic emergence. 

\subsection{Emergence and related notions}\label{ern}

In this Section, I give my preferred conceptions of theory, interpretation, and emergence. In Section \ref{emred}, I discuss the first condition for emergence, namely dependence, or linkage. In Section \ref{nnr}, I recall the notion of novel reference: which I will use in Section \ref{onep} to define ontological novelty.

Talking about emergence in science of course forces us to talk about theories: and so I will take a conception of theory, from De Haro (2016:~\S1.1) and De Haro and Butterfield~(2017:~\S2.2), that is appropriate for theories in the physical sciences.\\

{\it The conception of a `theory'.} The main task of applying a notion of theory to emergence is to make a conceptual distinction between the formalism and the interpretation of a theory. Thus we distinguish bare and interpreted theories:---

A {\it bare theory} is a triple $T:=\bra{\cal S},{\cal Q},{\cal D}\ket$ comprising a structured state space, ${\cal S}$, a structured set of quantities, ${\cal Q}$, and a dynamics, ${\cal D}$:\footnote{The notion of emergence given in this paper is of course not restricted to theories presented as triples. Any other appropriate formalisation of a bare theory will do, the main  alternative being that of a theory presented in terms of a (classical or quantum) partition function, and quantities derived from it. \label{triple}} together with a set of rules for evaluating physical quantities on the states.\footnote{In addition, there are {\it symmetries}. For simplicity of the exposition, I will avoid a systematic treatment of symmetries here (such a treament is given in De Haro and Butterfield (2017, 2019)), and consider them in the examples. Symmetries will resurface naturally in the example in Section \ref{mmpp}.}

An {\it interpreted theory} adds, to a bare theory, an {\it interpretation}: construed as a set (a triple) of partial maps, preserving appropriate structure, from the theory to the world. The interpretation fixes the reference of the terms in the theory. A bit more precisely, an intepretation maps the theory $T$ to a domain of application, $D$, within a (set of) possible world(s).\footnote{My use of possible worlds is not motivated by any version of modal realism, but is rather prompted by a basic fact of science: namely, that the ontologies of different scientific theories are often very different.} 
That is: there is a triple of maps, $i:T\rightarrow D$. Using different interpretation maps, the same theory can describe different domains of the world. In this paper, we will restrict attention to interpretations that are empirically adequate, so that they describe some significant domain of the world in sufficient detail.

This general conception of interpretation is logically weak, because little is required for a structure-preserving partial map. But to discuss ontological emergence (cf.~Section \ref{theoryem}), we need interpretations that are ``sufficiently good'' within their domain of application. Thus, in addition to the interpretation being empirically adequate, I will impose the additional condition that every element in the codomain is described by at least some element of the theory. Though this looks like a strong condition, it is in fact innocuous. The idea is captured by the requirement that the map be {\it surjective}: and, as you might expect, this can always be achieved by restricting the codomain. We will return to this notion of interpretation, and develop it, in Section \ref{nnr}. 

Discussions of emergence often work with {\it models,} i.e.~specific solutions of the dynamical equations of the theory that are physically permitted, rather than with entire theories. (The possibility of seeing emergence this way will resurface in the example of Section \ref{mmpp}.) In such cases, the interpretation maps are assigned to the states and quantities of a particular solution i.e.~model, rather than to the whole theory: and the domain is then naturally embedded in a single possible world.

A further refinement of the notion of interpretation that will aid better understanding of the notion of emergence is the distinction between the {\it intensions} and {\it extensions} of terms.\footnote{See Carnap (1947:~pp.~18-19), Lewis (1970), and Scheffler (1967:~pp.~54-57).} The extension of a term is its worldly reference under a certain interpretation, i.e.~the thing or entity being referred to, relative to a given possible world (with all of its contingent details). The intension is the linguistic meaning (cf.~the Fregean sense) of a term, as described by the theory. Thus although `Evening star' and `Morning star' have different intensions (i.e.~different linguistic meanings, and the stars they refer to are different in some possible worlds), they have the same extension in our world, viz.~the planet Venus---thus it is a contingent fact about our world that, while their intensions differ, their extension is the same. I will also use `extension' and `intension' for theories, as the ranges of the corresponding interpretations of the collection of all the terms of a theory.

Scientific theories have both intensions and extensions,\footnote{Cf.~Scheffler (1967:~pp.~54-57) and Nagel (1979:~pp.~95-113).} which in the present framework can be modelled by two different kinds of maps, each with its own domain, depending on whether the interpretation is an intension or an extension.\footnote{Lewis (1970:~pp.~22-27) defines intensions as maps from $n$-tuples of sequences of items---he calls such a sequence an `index'---to extensions, e.g.~the truth-values of sentences. And, as Lewis remarks, the framework also applies if the indices are construed as models consisting of states representing possible worlds (ibid, p.~23). Here, I simplify this situation by modelling both intensions and extensions by interpretation maps, rather than defining an extension as the (set of) objects/truth value(s), and an intension as a map to extensions.} Thus both intensions and extensions are structure-preserving partial maps from a bare theory or model to a domain relative to a possible world. The difference between the intension and the extension is in the kind of domain: explicitly, for states: an intension maps a state to a generic property (or physical arrangement) of a system mirroring the defining properties of the mapped state. Thus the image of the state and the domain abstract from contingencies such as how the system is spatially (and otherwise) related to other systems, and how the system is measured (so that the interpretation applies to all possible worlds that are described by the theory or model). By contrast, in the case of an extension, the image and domain are a fully concrete physical system: usually including also a specific context of experiment or description, and all the contingent details that are involved in applying a scientific theory to a concrete system.\footnote{For my earlier endorsements of intensional semantics, and more details, see De Haro and Butterfield (2017:~\S2.3) and De Haro and De Regt (2018:~\S1.1).}

I will construe the {\it domain of application,} $D$, introduced by the interpretation, as a set of {\it entities,} namely the elements of a set (fluids, particles, molecules, fields, charge properties, etc.), and relations between them (distances and correlation lengths, potentials and interaction strengths, etc.). Thus {\it the criterion of identity of domains is the set-theoretic criterion:} namely, the identity of the elements (and their relations). 

To explain a bit more how this criterion will be applied in practice, consider that we normally use a language to talk about the world: we describe the domain of application linguistically (e.g.~the theory may describe `this red ball I am kicking'), and express identity in terms of this language. This language is an aid (e.g.~words used to mention people) to describe on paper, or in sound, the intended physical things and properties, and it is in general different from the mathematical language of the bare theory. It is `the language of experimental physicists (and lay people)' used to describe the world, and it may contain extra-linguistic elements such as diagrams, images, ostension, etc. 
To give some examples of how this language picks out the objects in the world: (1)~~`A physicist's description of a laser beam' $\mapsto$ laser beam. (2)~~`A set of numbers on a computer screen' $\mapsto$ the events in a particle detector. (3)~~The descriptions of experimental results (including the description of the working of the instruments) that we find in the pages of scientific journals. 



Kuhn and Feyerabend said that the meanings of the terms of scientific theories cannot be compared, because theories are incommensurable. This cannot even be done empirically, because experimentation and observation are `theory laden'. Thus the Kuhn-Feyerabend critiques of meaning might lead one to think that the criterion of individuation of entities in the domain of application is problematic, if the language that we use to talk about them is theory-laden. I hope to address this issue in more detail elsewhere, but let me here make a comment of clarification in connection with the language which we use to talk about the domain of application.

For theories that, like those of the next Section, are related by linkage relations, the theory-ladenness of this language need not be an obstacle to comparing the domains of application, as my examples in Sections 5 and 6 will also illustrate: since the linkage relates the {\it bare theories} whose corresponding domains of application we compare. Therefore, combining linkage between bare theories, with the domain of application's reflecting the bare theory (its being `theory-laden'), and with the empirical data about the domain of application, we {\it can} compare the domains of application after all. Thus we do not need a putative ``theory-free'' language to compare them. Also, incommensurability is usually only {\it partial,} and the domains of application are often partly overlapping and partly distinct. 

Thus the practice of determining whether the elements of the domains of application of two theories are the same includes a combination of {\it theoretical} and {\it empirical} considerations. For examples of how to formalise and compare theories this way, see Sections \ref{mmpp} and \ref{casemm}.

I will expand on the criterion of identity of domains, in terms of intensions and extensions, in Section \ref{oee}. Let me now introduce the conception of emergence.

\subsubsection{The conception of emergence}\label{emred}

Consider the opening words of the {\it Stanford Encylopedia of Philosophy}'s entry on `Emergent Properties' (its only entry on `emergence'):

\begin{quote}\small
Emergence is a notorious philosophical term of art. A variety of theorists have appropriated it for their purposes ever since George Henry Lewes gave it a philosophical sense in his 1875 {\it Problems of Life and Mind.} We might roughly characterize the shared meaning thus: emergent entities (properties or substances) `arise' out of more fundamental entities and yet are {\it `novel' or `irreducible'} with respect to them... Each of the quoted terms is slippery in its own right, and their specifications yield the varied notions of emergence.\footnote{O'Connor and Wong (2002): the second Italics is mine. So far as I know, the broad idea of emergence goes back at least to 1843, when it was introduced by J.~S.~Mill as the idea that ``adding up the separate actions of the parts does not amount to the action of the living body'', even though he did not use the word `emergence'. Quoted in Landsman (2013:~pp.~379-380).}
\end{quote}
On this general characterisation, emergence is a two-place relation between two sets of entities: the emergent entities, and the more fundamental ones (I will dub them the `top' and `bottom' entities, respectively). The top entities are `novel' or `irreducible' compared to the bottom ones. 

`Irreducibility' here matches Broad's `lack of deducibility' quoted in the Introduction.\footnote{I already discussed how this notion seems problematic in the physical sciences, for reasons that Broad himself acknowledged: namely, it is too narrow, and there is a lack of convincing examples of such (strong) emergence.} As the italicized phrase in the above quote suggests, irreducibility and novelty are related: which naturally leads us to `novelty' as the weaker, more general alternative. For irreducibility is surely indicative of some type of novelty: if the top entities are irreducible compared to the bottom ones, then surely they have something novel. But the other way around is not the case: novelty need not be expressed as irreducibility. And so, I propose that `novelty', in this general and---admittedly---still unspecified sense, is the more general notion that one should seek to define emergence. Part of our task in developing a theory of emergence, then, is to further clarify what we mean by `novelty'.

The above discussion is echoed by the recent philosophy of physics literature, which sees `emergence' as a ``delicate balance'' between dependence, or rootedness, and independence, or autonomy (and the accounts also often compare theories rather than individual entities).\footnote{See e.g.~Humphreys (2016:~p.~26), Bedau (1997:~p.~375), Bedau and Humphreys (2008:~p.~1), Butterfield~(2011:~\S1.1.1).} Roughly speaking, dependence means that there is a {\it linkage} between two theories (or between two items of a given theory): while {\it in}dependence means that there is {\it novelty} in one theory with respect to the other (or between the two items of the given theory). This matches the two aspects---the two-place relation and the characterisation of the top entities as `novel'---of the above quote. 

In this Section, I shall make these notions more precise, within the framework for theories from Section \ref{ern}. I will take these two aspects---linkage and novelty---to align with the two aspects comprising a theory: viz.~the bare theory and its interpretation. Since these two aspects lie ``along different conceptual axes'', they can be happily reconciled: which makes it unnecessary to define novelty somehow as a `lack of linkage', `lack of deduction' or `irreducibility'. In this subsection, I discuss linkage, and in the next, novelty.\footnote{In his excellent, and very comprehensive, book, Humphreys (2016:~p.~26) distinguishes four features of emergence: dependence, novelty, autonomy, and holism. He only regards dependence and novelty as necessary for emergence, which are precisely the ones I consider here. For Humphreys' rationale for distinguishing novelty and autonomy, see (2016:~pp.~32-33).}

Thus I take {\it linkage} to be a {\it formal,} i.e.~uninterpreted, relation between bare theories (or between two items of a given bare theory), hence as an inter-theoretic relation between {\it bare theories}. More precisely, linkage is an asymmetric relation, $\mbox{\bf link}$, among the bare theories in a given family---with the family containing at least two theories. For the case of just two theories, it is a surjective and non-injective map, denoted as: $\mbox{\bf link}: T_{\sm b}\rightarrow T_{\sm t}$. Here I adopt Butterfield's mnemonic, whereby $T_{\sm b}$ stands for `best, bottom or basic', and $T_{\sm t}$ stands for `tainted, top or tangible', theory. Here, $T_{\sm b}$ normally denotes a single theory, but it can be used to denote a family of theories. So, the idea is that the bare theory, $T_{\sm b}$, contains some variable(s) which provide(s) a more accurate description of a given situation than does $T_{\sm t}$.\footnote{There are exceptions to this motivating idea, as we will see in the example of Section \ref{mmpp} (cf.~also the discussion in Section \ref{whichs}): where the bottom theory will be more explanatory, but the top theory will be more empirically accurate. What is essential is that the two theories are linked as here required.} (See the three cases of linkage, just below).

The idea of the linkage map, $\mbox{\bf link}$, is that it exhibits the bottom theory, $T_{\sm b}$, as approximated by the top theory, $T_{\sm t}$. The map's being surjective and non-injective embodies the idea of `coarse-graining to describe a physical situation' (but linkage is not restricted to mere coarse-graining: see below). The linkage map (and so, the broad meaning of `linkage', as an inter-theoretic relation, used here) can be any of three kinds which do not exclude one another, but are almost always found in combination:\\

(i)~~{\it A limit in the mathematical sense.} Some variable (e.g.~$V, c, \hbar\in\mathbb{R}_{\geq0}$ or $N\in\mathbb{N}$) of the theory $T_{\sm b}$ is taken to some special value (e.g.~$V\rightarrow\infty$, $N\rightarrow\infty$, $c\rightarrow\infty$, $\hbar\rightarrow0$), i.e.~there is a sequence in which a continuous or a discrete variable is taken to some value. $T_{\sm b}$ then refers to the sequence of theories obtained when the variable is taken to the limit, while $T_{\sm t}$ is the limit theory. 
In actual practice, the limit often comes with other operations, such as in (ii) immediately below---hence my warning above that these three kinds of linkage maps `are almost always found in combination'.\footnote{In the case of dimensionful parameters such as $c$ and $\hbar$, the limit is taken such that the parameter is large or small {\it relative to} other dimensionful parameters in the theory (such as the speeds of particles, in the case of $c$, and the value of energy differences and time intervals, in the case of $\hbar$). For a treatment of $c\rightarrow\infty$ limit in general relativity, see Malament (1986:~pp.~192-193). For a treatment of the $\hbar\rightarrow0$ case, see Landsman (2013:~pp.~380-383). See also my example in Section \ref{mmpp}.}

(ii)~~{\it Comparing different states or quantities.} Emergence often involves comparing a given physical situation, or system, or configuration of a system, to another that resembles it. This is often done by comparing some of the theory's {\it models}, i.e.~the solutions of the theory (for an example of this, see Section \ref{massmass}). Such physical comparisons have correlates in the formalism, which can be implemented prior to interpretation, by comparing {\it different states or quantities} or different {\it models}, here considered as uninterpreted solutions of the equations (cf.~Section \ref{ern}), between the bottom and top theories. In these cases, $T_{\sm b}$ is the bare theory that describes the situation (or system, or configuration) one wishes to compare to, and $T_{\sm t}$ is the bare theory that describes the situation (or system, or configuration) that one compares. The linkage map is then the formal implementation of this comparison, e.g.~the comparison between the formal, i.e.~uninterpreted, models.

(iii)~~{\it Mathematical approximations} (whether good or poor). Here, one compares theories (or expressions within theories) mathematically: perhaps numerically, or in terms of some parameter(s) of approximation. For example, as when one cuts off an infinite sequence beyond a given member of the sequence (perhaps because the sequence does not converge, or perhaps because there is no physical motivation for keeping an infinite number of members of the sequence); or when, in a Taylor series, one drops the terms after a certain order of interest. This will implement the idea that ``emergence need not require limits''.

Let me make three comments about how these three ways, (i)-(iii), of linking a theory to another, can define the map $\mbox{\bf link}:T_{\sm b}\rightarrow T_{\sm t}$. The first two comments are about how to apply (i)-(iii), while the third is about how linkage thus defined relates to emergence:

(1)~~Any of the three ways, (i)-(iii), of linking a theory to another, can define the map $\mbox{\bf link}:T_{\sm b}\rightarrow T_{\sm t}$. A given linkage map almost always involves a combination of several of the above approximations, as we will see in Sections \ref{imct} to \ref{casemm}. Also, the list is not meant to be exhaustive. 

(2)~~{\it Bridge principles:} before (i)-(iii) can successfully define a linkage map, one may need to ``choose the right variables'', or make additional assumptions. For example, one may need to first recast the bottom theory in a more convenient or more general form (see an example of this in Section \ref{bpp}). This goes under the name of {\it bridge principles} or bridge laws, i.e.~the map $\mbox{\bf link}:T_{\sm b}\rightarrow T_{\sm t}$ between the bottom and top theories may involve additional assumptions such as choices of initial or boundary conditions, or defining new variables to link the two theories' vocabularies. However, since the comparison here is between {\it bare theories,} the bridge principles here meant are formal, i.e.~uninterpreted, so that---recall the distinction at the end of Section \ref{ern}---the `language' just mentioned is the language of the bare theory, not the language we use to talk about the domain: thus it is mathematical rather than physical language.\footnote{Bridge laws or correspondence rules, which I here schematically dub `bridge principles' (to avoid the physical connotations of `correspondence' and `laws') are discussed in Nagel (1949:~p.~302) and (1961:~p.~354). The latter reference discusses two roles (as `conditions of connectability' and `derivability') that bridge principles fulfill, and gives three types of bridge laws. My bridge principles here are more restricted, in that they connect formal, i.e.~uninterpreted theories, as explained in the main text. For a discussion of ``ontological reduction'', see the end of Section \ref{oee}, especially footnote \ref{ontred2}}

(3)~~The ways of linking, (i)-(iii), as here defined are formal i.e.~not interpretative, as I stressed in (ii): they link the bare theories, but they are physically motivated, and they often have a correlate in the theory's interpretation (which serves as a constraining affordance on the kind of linkage map one is likely to adopt). Just {\it how} the theory's interpretation correlates with the formal linkage map is in fact the very question of emergence itself, as I will discuss in the next few Sections:
\\

{\it Emergence}. We have emergence iff two bare theories, $T_{\sm b}$ and $T_{\sm t}$, are related by a linkage map, and if {\it in addition} the interpreted top theory has novel aspects relative to the interpreted bottom theory.\footnote{This definition can also be adapted for the emergence of properties, entities, or behaviour within a single theory. I will use this in the example in Section \ref{bpp}.}

The linkage map thus specifies the dependence part of the emergence relation. To characterise emergence, we still need to specify the novelty.\footnote{A general characterisation of novelty, as `not being included in (the closure of) a domain', is given in Humphreys (2016:~p.~26). Butterfield (2011:~\S1.1.1) adds `robustness' (i.e.~the top theory's independence from the bottom theory's details) as an {\it additional requirement} for emergence. While I agree that robustness is an important property to further analyse the {\it physical significance} of emergent behaviour, and I will implement it in my choice of the {\it emergence base} (i.e.~the class of bottom theories with respect to which the top theory emerges) in Eqs.~\eq{parx0}-\eq{parx} below: I do not think robustness is a necessary requirement for a minimal ontological description of emergence: see footnote \ref{basepoint} for other choices of emergence base. \label{robust}} 

Philosophers often distinguish between ontological and epistemic emergence. The intended contrast is, roughly, between  emergence `in the world' vs.~emergence merely in our description of the world. Thus according to this distinction we can have two types of {\it novelty}: ontological or epistemic. Specifying novelty in the case of ontological emergence is the topic of the next Section.

\subsubsection{Ontological emergence as novel reference}\label{nnr}

In this Section, I characterise the kind of novelty that is relevant to ontological emergence. First of all, let me point to an obvious question that comes to mind about the literature on ontological emergence. We saw, in Section \ref{emred}, that it is reasonable to define emergence in terms of novelty: this being a more general notion than the stronger `irreducibility'---and, as I mentioned, this also seems to reflect a consensus in the recent philosophy of physics literature. Thus one would expect that, when defining {\it ontological} emergence, one would try to further characterise the so-far unspecified notion of `novelty' in terms that are relevant to the ontology of the entities or theories involved in a relation of emergence. 

But the literature usually takes some highly {\it specific} metaphysical relation instead, whose relation to novelty is not explicitly addressed, and uses that to characterise the emergence relation between the top and bottom entities or theories. I think this shift to specificity is understandable taking into account some of metaphysicians' interests, but I will also argue that, lacking a general notion of ontological {\it novelty,} we run the risk of missing some more basic aspects of what we mean by `ontological emergence'.

For example, Wong (2010:~p.~7) defines ontological emergence as `aggregativity generating emergent properties': `Ontological emergence is the thesis that when aggregates of microphysical properties attain a requisite level of complexity, they generate and (perhaps) sustain emergent natural properties.' And Wimsatt (1997, 2007:~Chapter 12) takes the {\it lack of aggregativity} of a compound as the mark of emergence.\footnote{I take `lack of aggregativity' to be an example of ontological rather than epistemic emergence, since aggregativity is a property of entities rather than theories.}

Further, in the {\it Stanford Encyclopedia of Philosophy} article quoted earlier, O'Connor and Wong (2002:~\S3.2) review various other uses of the phrase `ontological emergence', none of which are obviously introduced as an explication of `novelty': viz.~ontological emergence as `supervenience', as `non-synchronic and causal', and as `fusion'.

I will not criticise these accounts here, since my claim is {\it not} that they are incorrect,\footnote{For a criticism of the supervenience account, see Butterfield (2011:~pp.~948-956).} but rather that the accounts---interesting as they are---are very {\it specific:} in fact, too specific to be able deal with all the examples that physicists are interested in. For example, Wimsatt's (1997, 2007) `failure of aggregativity' does not seem applicable in cases where one compares systems that are neither spatial components of each other, nor aggregates of components. In other words, the relevant relation between the levels is not always one of spatial inclusion or material constitution. Thus already my examples from Sections \ref{imct} and \ref{mmpp} are not covered by Wimsatt's criterion of failure of aggregativity. Also, it would seem that failure of aggregativity is most interesting {\it if it leads to novelty.} The same can be said about causation: it may play a role in important examples of emergence such as the mind, but it is hard to see how it plays a role in the kind of examples that I discuss in this paper, which are, I believe, ``garden variety'' for the theoretical physicist, and involve {\it no causation} from the bottom to the top entities. 

Another reason to be sceptical about too specific accounts of ontological novelty is the requirement of consistency with the general description of emergence reviewed in Section \ref{emred}. As I said above, having settled for {\it novelty} as the general mark of emergence, one naturally expects an explication of emergence to point to novelty in the top theory's ontology. Thus I take it that the metaphysical accounts of emergence just discussed, in so far as they all capture aspects of emergence, aim to put forward a specific metaphysical expression of `novelty'. This is apparent from Wong's mention of the `generation of emergent natural properties': surely what makes these properties of the top theory `emergent' (on pain of his own definition being circular) is that they are novel, relative to the properties of the bottom theory.  


My account of ontological emergence is metaphysically pluralist in that, while it recognises that emergence is a matter of {\it novelty} in the world (as the general notion of emergence prompts us to say), it does not point to a {\it single} metaphysical relation (e.g.~supervenience, causal influence or fusion) as constitutive of ontological emergence. And I believe that this is just right, for the reasons above: a single metaphysical relation does not seem to cover all the cases of interest. Indeed my main contention is that there is a more basic framework for `ontological novelty' that needs to be developed before one fruitfully moves on to detailed metaphysical analyses. Furthermore, I take it as a virtue of the present framework that it allows for an explication of the phrase `ontological emergence' without appealing to specific, and sometimes controversial, metaphysical notions. So it seems that an account satisfying these desiderata should exist (more on this in Section \ref{mando}). In this sense, my account is ``basic'', and can perhaps best be seen as an attempt to state the ``almost obvious'' in a more precise way. \\
\\
To discuss ontological novelty, then, I take my cue from Norton (2012), which is a perceptive discussion of the difference between {\it idealisation} and---what I shall call---{\it non-idealising approximation.} Roughly speaking, he proposes the following contrast:

\begin{quote}\small A [{\it non-idealising}] {\it approximation} is an inexact description of a target system.\footnote{Norton calls this an ``approximation'', but this I already used this word in Section \ref{emred} (iii) in the more common sense of a `mathematical approximation', hence my addition of `non-idealising' here. For it seems to me that doing an approximation does not, in general, prevent the approximation from being an idealisation, i.e.~an idealising system could exist (and this is also the jargon adopted in the literature on reduction, see e.g.~Schaffner (1967:~p.~144) and Nagel (1979)). Therefore, I dub the more restrictive type (which does not play an important role in the rest of the paper) a `non-idealising approximation'. \label{nia}} 
An {\it idealization} is a real or fictitious, idealizing system, [possibly] distinct from the target system, whose properties provide an [exact or] inexact description of the target system' (p.~209).\footnote{Norton's use of the gerundive `idealizing', here in his definition of `idealization', is not circular, because `idealizing', introduced as an adjective for `system', is merely a label used to distinguish this system from the other system involved: namely, the `target' system. See my use of these words in Section \ref{whichs}.}
\end{quote}
Norton summarises the main difference between the two as an answer to the question: `Do the terms involve novel reference?' On Norton's usage, only idealisations introduce reference to a novel system---notice that the word `system' here may refer to a real or a fictitious system. In the case of idealisation, there is an `idealised system' that realizes the `idealised properties'. In the case of a non-idealising approximation, an idealised {\it system} either does not exist, or is not accurate enough to describe the target system under study.\footnote{An idealising system is sometimes called a `limit system', the word `limit' here echoing case (i) in Section \ref{emred}. However, I deny that idealisations invariably require limits, because they only require a physical system described by a top theory (more on this in Section \ref{nornc}), and thus all of (i)-(iii) can correlate with an idealisation. For example, the case study in Sections \ref{bpp}-\ref{encai}, although naturally written as a limit, does not require one since the theory (as opposed to some of its models) contains no singularities. See the discussion in Section \ref{swe}. For this reason, I do not use the `limit system' jargon.}

I propose to define ontological novelty in terms of novel reference thus construed. Thus, this gives an additional condition on the linkage map defined in Section \ref{emred}: namely, {\it the composition of the linkage map and the top theory's interpretation map, $i_{\sm t}\,\circ\,\mbox{\bf link}$, must have an idealisation in its range}, in the sense that the top theory has a referent that is novel, relative to the bottom theory's domain of application.\footnote{Humphreys (2016:~p.~29) considers novelty within a single domain, while I here require novelty of one domain relative to another. The domain $D_{\tn t}$ is novel, relative to $D_{\tn b}$, if some of its elements or relations are not included in $D_{\tn b}$. For more details, see Section \ref{nornc}. For novelty of {\it entities}, see Humphreys (2016:~p.~34).} 

This is not {\it merely} the condition that the top theory must describe a physical system through its interpretation map, $i_{\sm t}$: for the top theory is linked to the bottom theory, and so this is an extra condition on the linkage map between the two theories. 

Although this characterisation of ontological novelty as novelty of reference is, by itself, not a new construal of the notion (see the discussion in Section \ref{comparrw}), it will give an interesting criterion for {\it emergence}: one formalised in terms of maps, in Section \ref{onep}. 

There is `novel reference' when the terms of the bare theory refer (via the interpretation maps introduced in Section \ref{emred}) to new things in the world, i.e.~elements or relations that are (a) not in the domain of application of the bottom theory, $T_{\sm b}$, that $T_{\sm t}$ is being compared to, but are (b) still in the world---they are in the domain of application of $T_{\sm t}$. 

Novelty is here not meant primarily in the temporal sense, since it is conceptual novelty\footnote{Humphreys (2016:~p.~39) distinguishes ontological from conceptual novelty. His `conceptual novelty' corresponds, roughly, to what I call `epistemic emergence' (see Section \ref{nornc}). His `ontological novelty' does not exactly match mine, which contains aspects of his ontological and his conceptual novelty. See the discussion in Section \ref{comparrw}. By `conceptual' I here mean the qualitative notions involved in interpreting a theory (e.g.~the notion of `simultaneity' in classical mechanics as a relation between events in the world), rather than the uninterpreted notions of the bare theory. This is why `conceptual', on my construal, pertains to the interpretation of the theory, and hence to its ontology, rather than to its descriptive apparatus.} that counts (cf.~Nagel (1961:~p.~375)),\footnote{Of course, the linkage map does not exclude the temporal dimension: `diachonic' emergence is obtained if the linkage map entails ``evolving the time parameter''. However, this will not be my focus, since it is a special case of the general linkage map. For interesting treatments of diachronic emergence, see Guay and Sartenaer (2016:~pp.~301-302, 316-317) and Humphreys (2016:~pp.~43-44, 66, 70-72).} even allowing reference to other possible worlds (as Norton does when he refers to idealisations as referring to real or fictitious systems:~p.~209; cf.~also the definition of an interpreted theory in Section \ref{ern}). 

For example, Norton considers the eighteenth- and nineteenth-century theories of heat to be referentially successful, despite the fact that we now know that heat is not a fluid. According to Norton, the theory is nevertheless---
\begin{quote}\small
referentially successful in that the idealizing system is a part of the same system the successor theory describes. The ``caloric'' of caloric theory refers to the same thing as the ``heat'' of thermodynamics, but in the confines of situations in which there is no interchange of heat and work.
\end{quote}

I propose to read Norton's statement above as saying that the theory is referentially successful because it refers to an idealised world\footnote{Norton's suggestion that novel reference characterises idealisation seems to rule out purely {\it instrumentalist} interpretations of theories: but only in so far as such interpretations can successfully avoid ontological questions (which is far from being an innocuous condition!). Thus from now on, the interpretations considered will not include strongly instrumentalist interpretations of theories. Other philosophical positions, such as empiricism, are of course supported by the framework: see footnote \ref{empiricism}. For a critique of instrumentalism in physics, see Wallace (2012:~pp.~24-28).\label{instrumentalism}}---a possible world in which there is no interchange of heat and work---that approximates the target system of interest well, according to standard criteria of empirical adequacy: and that as such it is referentially successful. I develop this suggestion in the next Section.

The explication of ontological emergence as novel reference does not attempt to answer questions about scientific realism or referential stability across theory change.\footnote{For a summary of the debate, see Radder (2012:~\S3.4), Psillos (1999:~pp.~282-284). The example of phlogiston is discussed in Kitcher (1978:~pp.~529-536), and water is in Chang (2012:~\S4.1, \S4.3.4).}
In particular, $T_{\sm b}$ should not strictly be seen as the successor of $T_{\sm t}$ in a historical sense, nor need the continuity between their domains of applicability be assumed.

\subsection{Ontological emergence, in more detail}\label{onep}

In this Section, I propose an answer to the question of how to characterise ontological emergence for theories formulated as in Section \ref{ern}. My proposal is based on a restatement of the notion of `novel reference': namely as the failure of the interpretative and linkage map to mesh, i.e.~to commute, in the usual mathematical sense of their diagram's ``not closing''. This idea is inspired by Butterfield (2011a:~\S3.3.2), though I here consider it for {\it interpretations} rather than for bare theories: a distinction which will turn out to be important (cf.~Section \ref{oee}).

I begin with two remarks which further specify the context to which these concepts apply:\\

(1) {\it Emergence of theories: and of models, entities, properties, or behaviour.} I will formulate the distinction in terms of {\it theories}, but the idea can be equally well formulated in terms of models (cf.~Section \ref{ern}), entities, properties, or behaviour (and I will adopt the latter perspective in the examples in Sections \ref{imct} and \ref{mmpp}). 

(2) {\it The space of theories.} Properly defining a linkage map, especially if it involves a mathematical limit of a theory (so that one can then discuss the sequence of corresponding interpretations), often requires introducing the notion of a space of theories related by the linkage map. Sections \ref{mmpp} and \ref{casemm} will take a simpler approach and consider sequences of states, quantities, and dynamics, without explicitly specifying the full details of the spaces in which the sequences are defined, so as to simplify the presentation.\footnote{Thus I refrain from a more ambitious technical aim, of trying to define general conditions of emergence (and, especially, the aim of characterising the robustness of the emergent behaviour---which, as I said in footnote \ref{robust}, I do not think is required for emergence) in terms of the underlying topology, and choice of it. For an example of deploying topology in the ``space of theories'', cf.~Fletcher (2016).}\\

Section \ref{nornc} contains the core of this Section and the main proposal of this paper, i.e.~the explication of ontological  emergence. In Section \ref{oee}, I discuss to what extent the proposal answers the questions in the Introduction. 

\subsubsection{Epistemic and ontological emergence as commutativity and non-commu\-ta\-tivity of maps}\label{nornc}

In this Section, I reformulate the notion of novel reference in terms of the non-commutati\-vity of the interpretative and linkage maps, or their ``failure to mesh'': which I will use to give an explication of ontological emergence. I here mean non-commutativity of maps in the usual mathematical sense that, depending of the order in which the maps are applied, the image (value, output) that a given argument (input) is mapped to varies---the mismatch thus expressing the {\it novelty} of the reference.

Thus the starting point is a reformulation of the idea of novel reference, in terms of my notion of interpretation of a bare theory (Section \ref{ern}). Recall that an interpretation is a surjective map, $i$, from the bare theory, $T$, to a domain of application, $D$, $i:T\rightarrow D$, preserving appropriate structures. Norton's idea can then be formulated as follows.

Consider two bare theories, a bottom theory, $T_{\sm b}$, and a top theory, $T_{\sm t}$. Their interpretations are corresponding surjective, structure-preserving maps (by our assumption, in the preamble of Section \ref{ern}, that the interpretation is ``sufficiently good''), from the theories to their respective domains of application. That is, there are maps
$i_{\sm b}:T_{\sm b}\rightarrow D_{\sm b}$ and $i_{\sm t}:T_{\sm t}\rightarrow D_{\sm t}$. 

Now, consider, as in Section \ref{emred}, a linkage map, $\mbox{\bf link}:T_{\sm b}\rightarrow T_{\sm t}$, mapping the bottom theory to the top theory. 
There are two interpretative cases,  as follows. 

\begin{figure}
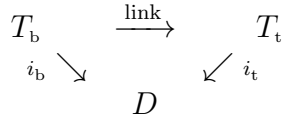

\begin{center}
\bea
\begin{array}{ccc}T_{\tn b}\!\!&\xrightarrow{\makebox[.6cm]{$\sm{link}$}}&\!T_{\tn t}\\
~~~~~~~{\sm{$i$}}_{\tn{b}}\searrow&&\swarrow \sm{$i$}_{\tn{t}}~~~~~~~~\\
&D&\end{array}\nonumber
\eea
\caption{\small Possibility of epistemic emergence. The two interpretations describe ``the same sector of reality'', so that $i_{\sm b}=i_{\sm t}\circ\,\mbox{link}$.}
\label{3leginterp}
\end{center}
\end{figure}

\paragraph{(1)~~Allowing epistemic emergence.} If the two theories describe the same ``sector of reality'', i.e.~the same domain of application in the world, then the ranges of their interpretations must coincide. (Alternatively, the top range must be contained in the bottom range, i.e.~$D_{\sm t}\subseteq D_{\sm b}$: a condition that I will discuss separately, below). For the range of the interpretation is what the theory purports to represent: and  these two theories represent the same sector of reality. This situation of identity is described by the commuting diagram in Figure \ref{3leginterp}. As is clear from the diagram, there is no novel reference here, because the domains of application in the world are the same. 

Notice that, since the domains of application are the same, the two maps map to the same elements within that domain of application (the same experiments, interactions, correlations, etc.): even though, as interpretations, they are of course different, because they map from different theories (and so, one theory could describe water in terms of molecular dynamics of the single molecules, or in terms of the dynamics of groups of molecules, even though both describe the same empirical data). In such a case, the two interpretations are related as follows: $i_{\sm b}=i_{\sm t}\,\circ\,\mbox{\bf link}$. 

Another type of epistemic emergence is if the bare theories, $T_{\sm b}$ and $T_{\sm t}$, are equivalent bare theories. That is, we consider the commuting diagram in Figure \ref{3leginterp}, but weaken the linkage map {\bf link}, allowing it to be injective. There is no novelty in the world, but only in the bare theory and the interpretation map (but not in its range).\footnote{Another interesting case is equivalent theories with {\it different} interpretations. Though there is ontological novelty here, there is no ontological emergence because the linkage map is injective: see van Dongen et al.~(2019:~footnote 52).} Dualities in physics give vivid examples of this type of emergence:\footnote{As discussed in van Dongen et al.~(2019:~Section 4.4),  De Haro (2015:~p.~118), and Dieks et al.~(2015:~pp.~208-209), dualities of this type only give cases of epistemic emergence. The reason is that the linkage map between the two theories is an isomorphism, which is in tension with the asymmetry (non-injectivity) required for ontological emergence.} see Castellani and De Haro (2019) (see also Section \ref{swe}).

Since in Figure \ref{3leginterp} there is no novel reference, there cannot be ontological emergence. However, there can be {\it epistemic emergence}: this will be the case when the two interpreted theories describe the domain of application differently. And so, the top theory $T_{\sm t}$ can have properties or descriptive {\it features} that are indeed novel, and striking, relative to the bottom theory, $T_{\sm b}$. The novelty of these properties and descriptive features, however, is not a matter of {\it reference}: it does not rely on a difference in elements or relations between the bottom and top domains of application, since by definition we have a commuting diagram in Figure \ref{3leginterp}. Rather, the novelty lies in the different relations between the bare theory and the domain that $i_{\sm t}$ establishes, compared to $i_{\sm b}$, i.e.~in the kind of statements that the interpreted theory makes about the domains of application.

I now discuss the special case of proper inclusion, i.e.~$D_{\sm t}\subset D_{\sm b}$. The reason for allowing this case is that there are surely cases of epistemic emergence in which the diagram ``does not close''. If $i_{\sm b}$ is a ``very fine-grained interpretation'', ``respecting'' the distinction of the many micro-variables of $T_{\sm b}$, i.e.~an injective map, while link is of course non-injective, then we have a diagram which cannot possibly commute on all arguments. But this could be a case of mere coarse-graining with no ontological novelty, while the failure of the diagram to close would seem to suggest ontological emergence. In such cases, empirical adequacy (see Section \ref{ern}) on the domain $D_{\sm b}$, so that it is a good range for $i_{\sm b}$, requires $D_{\sm b}$ to ``contain coarse-grained facts'', since $i_{\sm b}$ needs to be surjective on $D_{\sm b}$. Thus in such cases, ``good'' interpretations of $T_{\sm b}$ include a classification of microstates into macrostates. 

To sum up: cases where the map $\mbox{\bf link}$ is {\it mere coarse-graining}, which results in the ``dropping of certain fine-grained facts from the domain'', do not count as ontological emergence because then $D_{\sm t}$ is a proper subset of $D_{\sm b}$. \\

Let me spell out the possibility of epistemic emergence a bit more. Though the domains and ranges of $i_{\sm b}$ and $i_{\sm t}\,\circ\,\mbox{\bf link}$ are the same, the interpretative maps $i_{\sm b}$ and $i_{\sm t}$ are different and have different theories as their domain of definition. Thus the novelty lies in the existence of a new theory, $T_{\sm t}$, with its own novel interpretation map, $i_{\sm t}$; and this novelty is relative to the bottom theory, $T_{\sm b}$, with its interpretation map $i_{\sm b}$. Therefore the emergence is epistemic, since it takes place at the level of the theory and its interpretation map (while the domains are the same): it is emergence in the theoretical description, rather than emergence in what the theory describes.\footnote{
Epistemic emergence is a matter of novel description by the bare theory, and so an anonymous referee asked: `How do we determine what a bare theory describes, prior to interpretation?' The answer is that we do not need to determine epistemic emergence in complete independence of the interpretation map. However, a bare theory, even if physically uninterpreted, already comes with a mathematical physics language that allows us to see the novelty in the bare theory: it is the linkage function itself that introduces this epistemic novelty (see Section \ref{ern} and De Haro (2016:~Section 1.1.1; 2019:~Section 2.1.1)). In the case of dual theories with the same domain of application, the two theories give different effective descriptions of this domain, because of the approximations introduced: for details, see Castellani and De Haro (2019:~Section 2.2). In both cases, the novelty is in both the bare theory and in the interpretation map, but---crucially---the interpretation map's range is the same, and the diagram in Figure \ref{3leginterp} closes: and so, the distinction with the case of ontological emergence, to be made below, {\it is} sharp. For explicit examples of how one formalises a theory in this way, see Sections \ref{mmpp} and \ref{casemm}.}

Epistemic emergence should not be dismissed as ``merely'' a matter of ``words'', or even ``taste''. For theories, and interpretations of the kind here discussed, are not the kinds of things that one can choose as one pleases---there may be virtually no freedom in choosing an appropriate linkage map giving rise to a {\it consistent} theory $T_{\sm t}$ whose interpretation is {\it as good} as that of $T_{\sm b}$---i.e.~the ranges of their interpretations {\it must be} identical (cf.~Section \ref{oee}).

The above is, as it should be, only a {\it necessary condition of epistemic emergence}: as a difference in theoretical description. To give a sufficient condition for epistemic emergence, one needs a more precise characterisation of epistemic novelty. Epistemic novelty is often characterised in terms of computational or algorithmical complexity, chaotic behaviour, possibility of derivation only through simulation, etc. Alternatively, it is seen as the failure of prediction or of explanation.\footnote{For more details, see, for example, Bedau (1997:~pp.~378-379), Bedau and Humphreys (2008:~p.~16), Humphreys (2016:~pp.~144-197).} But since epistemic emergence will not be my main topic, I will leave such further characterisations aside (I will return to it in the example of Section \ref{swe}; see also Castellani and De Haro (2019:~\S2.3)). \\

\paragraph{(2) Non-commutation: ontological emergence.} But when there is {\it novel reference}, the two interpretations refer to different things---they may even refer to different domains of application. For example, hydrodynamics describes the motion of water between $0^{\sm o}$ C and $99^{\sm o}$ C at typical pressures, while the molecular theory of water potentially describes it in a different domain of application, since it can e.g.~explain the boiling of water as the breaking of the inter-molecular hydrogen bonds due to molecular excitations. Here, one should not confuse the fact that the interpretations of the two theories attempt to explain the properties of the same target system (e.g.~turbulence in a fluid) with the distinctiveness of their ranges: each of which is subject to its own conditions of empirical adequacy, within its relevant accuracy.

In all such cases of novel reference, the domains of application are different: 
\bea\label{dnotd}
D_{\sm t}\nsubseteq D_{\sm b}~,
\eea
and therefore so are the interpretation maps, that is:
\bea\label{itnotib}
i_{\sm t}\circ\,\mbox{link}&\not=&i_{\sm b}\\
\mbox{ran}(i_{\sm t})\!&~/\!\!\!\!\!\!\subset&\mbox{ran}(i_{\sm b})~.\nonumber
\eea
Thus the range of the  interpretation of the top theory, $T_{\sm t}$, is not the same as the range of the interpretation of the bottom theory, $T_{\sm b}$, nor is it a subset of it. This is what I call {\it ontological emergence}, and it is pictured in Figure \ref{4leginterp}.\\

There are cases in which $T_{\sm t}$ is not a good approximation to a single theory $T_{\sm b}$ (cf.~(i) and (iii) in Section \ref{emred}), but only to one of the members of a {\it family} of theories $T_{\sm b}$. Let us label this family by $x$, so that the bottom theory becomes a function of $x$, which we write as $T_{\sm b}(x)$. In other words, we take the emergence base (i.e.~the class of bottom theories with respect to which the top theory emerges: cf.~footnote \ref{robust}) to be the {\it entire family} of theories, $\{T_{\sm b}(x)\}$, for any $x$ in an appropriately chosen range. We will see this at work in the example of Section \ref{mmpp}.

Since the emergence base is the entire set of theories, $\{T_{\sm b}(x)\}$, the bottom theories all have the same interpretation {\it map}, $i_{\sm b}$, but the {\it range} of $i_{\sm b}$ now depends on, i.e.~varies with, $x$ through the bottom theory, i.e.~we evaluate $i_{\sm b}(T_{\sm b}(x))$. 

Now take $x$ to be bounded from below by $0$ and consider, as a {\it special case,} a linkage map that is taking the limit $x\rightarrow0$, i.e.:
\bea\label{parx0}
T_{\sm t}:=\lim_{x\rightarrow0}T_{\sm b}(x)~.
\eea
In this case, the statement of ontological emergence, Eqs.~\eq{dnotd}-\eq{itnotib}, amounts to:
\bea\label{rani}
\mbox{ran}(i_{\sm t})\nsubseteq\mbox{ran}\,(i_{\sm b})|_{x=0}
\eea
or, perhaps more intuitively:\footnote{The condition $x=0$ in the subscripts on the right-hand side is motivated by the condition, mentioned above, that we take the {\it entire class} of theories, $\{T_{\tn b}(x)\}$, to belong to the {\it emergence base}. This will be important in footnote \ref{curiet}. But there are other possible choices of this base. In particular, by analogy with the case of just two theories, it would suffice that $D_{\tn t}\not=D_{\tn b}(x_0)$, where $x_0$ is some reference value of $x$, for us to claim that the top theory emerges from the bottom theory (where the bottom theory is here identified with $T_{\tn b}(x_0)$). However, it is in most cases better motivated to regard the whole class of theories, for any $x$, as the emergence base, i.e.~to take $\{T_b(x)\}_{0\leq x\leq x_0}$ as the emergence base, under the assumption that $\mbox{ran}\,(i_{\tn b}(x))=\mbox{ran}\,(i_{\tn b}(x'))$ (where $i_{\tn b}(x)$ is $T_{\tn b}(x)$'s own interpretation map) whenever $x,x'\in[0,x_0]$. For then $T_{\tn t}$ is novel relative to the whole class of theories $\{T_b(x)\}_{0\leq x\leq x_0}$, all of which have the same interpretation, $i_{\tn b}$. It is {\it this} stronger requirement that I follow in this paper, and it amounts to a {\it robustness} condition.  The requirement can be easily weakened as needed, by evaluating the right-hand side of Eq.~\eq{rani} at some other point $x$, or by adapting the emergence base in a way appropriate for the situation at hand. Also Humphreys (2016:~p.~28) emphasises the relativity of `emergence' to the emergence base: my choice $x=0$, on the right-hand side of Eq.~\eq{rani}, is how my framework reflects this. Additional motivation for this choice is as follows. Assume that there were some $x'\in(0,x_0)$, such that $\mbox{ran}\,(i_{\tn b}(x))\not=\mbox{ran}\,(i_{\tn b}(x'))$. Then the theory at $x=x'$ is already emergent: in other words, ``we did not need to go to $x=0$ to get emergence''. So, it is best to take the emergent theory, $T_{\tn t}$, to be the one whose domain is distinct from the domains of all the theories at $x\in[0,x_0]$ (alternatively, to define $x$ such that $D_{\tn t}$ is distinct from $D_{\tn b}(x)$ for the entire range of $x$).\label{basepoint}}
\bea\label{parx}
D_{\sm t}:=i_{\sm t}(T_{\sm t})\nsubseteq D_{\sm b}|_{x=0}:=i_{\sm b}(T_{\sm b}(x))|_{x=0}~.
\eea
In other words, the domains that we get by interpreting the top theory directly, or by first interpreting the bottom theory and then setting $x=0$, are different, as we will see in the example below.\footnote{This also applies to cases where, instead of a continuous parameter $x$, we have a sequence of bottom theories, $T_{\tn b}, T_{\tn b}',T_{\tn b}'',\ldots$, and a sequence of interpretations $i_{\tn b}, i_{\tn b}',i_{\tn b}'',\ldots$ In such cases, ontological emergence lies in the sequence of interpretative maps not converging to, or being distinct from, the given approximation (cf.~(i) and (iii) in Section \ref{emred}), the top theory's interpretative map, $i_{\tn t}$.}

\begin{figure}
\begin{center}
\bea
\begin{array}{c}
{\sm{link}}\\
\end{array}\!\!\!\!\!\!\!\!\!\!\!
\begin{array}{ccc}T_{\sm t}&\xrightarrow{\makebox[.6cm]{$\sm{$i_{\tn t}$}$}}&D_{\sm t}\\
~~~~\big\uparrow~~~~&&
\rotatebox[origin=c]{-90}{$\not=$}\\
T_{\sm b}&\xrightarrow{\makebox[.6cm]{$\sm{$i_{\tn b}$}$}}&D_{\sm b}
\end{array}\nonumber
\eea
\caption{\small The failure of interpretation and linkage to commute ($i_{\sm b}\not=i_{\sm t}\circ\,\mbox{link}$) gives rise to different interpretations, possibly with different domains of application, $D_{\sm b}\not=D_{\sm t}$.}
\label{4leginterp}
\end{center}
\end{figure}

In view of the difference between Figures \ref{3leginterp} and \ref{4leginterp}, we can now reformulate novel reference as {\it the linkage map's failure to commute, or to mesh, with the interpretation}. 
When linkage and interpretation do not commute, the two interpretation maps are different because the theories refer to different domains, and different systems---even if the underlying physical object, like the sample of water above, may of course be the same.

It is possible to formulate a {\it robustness} condition for the emergent behaviour---which I will assume in the rest of the paper---as the condition that the emergence base is a whole class of theories, i.e.~$\{T_b(x)\}$ for a range of values of $x$, rather than a single theory for fixed $x$. Thus, emergence is robust if the interpretation does not change as we vary $x$ over the base (see footnote \ref{basepoint}).

To sum up: the meshing condition between the linkage and interpretation maps, formulated as the lack of commutativity or of closing of their joint diagram, Figure \ref{4leginterp}, should be seen as a reformulation of novelty of reference---a reformulation that gives a straightforward formal criterion that can be used in examples. 

Let me briefly discuss the relation between reduction and emergence. I will endorse the philosopher's traditional account of reduction: namely, Nagel's view\footnote{The philosopher's traditional account of reduction is: Nagel (1961:~pp.~351-363; 1979), Hempel (1966:~\S8), Schaffner (2012). For rebuttals of the objections to this traditional account, see Dizadji-Bahmani et al.~(2010:~pp.~403-410) and Butterfield (2011a:~\S3).}---as, essentially, deduction of one (here, bare) theory from another, almost always using additional definitions or bridge-principles linking the two theories' vocabularies.\footnote{By the `vocabulary' of a theory, I here mean the mathematical rules and the words of ordinary or technical language used to express propositions within the theory, as well as its rules of inference.\label{vocab}} Not all linkage maps will be reductive. For the relations between physical theories are often much more varied than logical deduction with bridge principles allows (for example, the use of limits and related procedures, such as renormalization, often adds content to the theory). But when the linkage {\it is} reductive, the account explains at once {\it why} ontological emergence and reduction are independent of each other: for the former is a property of the theory's {\it interpretation} (i.e.~the novelty in the domain), while the latter, understood as a formal relation, is a property of the linkage map only, i.e.~a relation between {\it bare theories}.\footnote{Notice that the formal reduction between bare theories discussed here contrasts with the more controversial notion of ``ontological reduction''. For example, Hendry (2010:~pp.~184, 188) distinguishes between inter-theoretic reduction (which is what philosophers have traditionally meant by reduction: see the main text) and ontological reduction. The former sense will be my main meaning of `reduction': and so, when I say that ``emergence and reduction are compatible'', I mean reduction in this sense. See also the end of Section \ref{oee}, especially footnote \ref{ontred2}.\label{ontred}}

\subsubsection{Is emergence ubiquitous? Regimenting the uses}\label{oee}

In this Section, I point out two properties of the regimentation of `emergence' I have proposed, that have the effect of reducing the number of putative cases of emergence; and I analyse the extent to which the framework gives a clear-cut criterion of emergence.

Understanding ontological emergence as novelty of reference, and epistemic emergence as novelty of theoretical description, gives a helpful regimentation of the uses of the word `emergence'. One problem which seems to have plagued the literature is the apparent ubiquitousness of emergence, which would seem to be sanctioned by my logically weak conception in Section \ref{ern} (cf.~Chalmers (2006:~\S3)). 

In addressing this problem, the first point to note is that, as I stressed: given the wealth of examples in which physicists justifiably talk about emergence (and which the philosophical literature has also endorsed), pervasiveness is something one may, to some extent, expect and accept.\footnote{I agree with Humphreys (2016:~pp.~54-55), who criticises the `rarity heuristic', i.e.~the idea that a construal of emergence must make it a rare property in order for it to be correct. Ontological emergence may well be pervasive: what I aim at here is clarifying how it is constrained. See also Bedau and Humphreys (2008:~pp.~12-13).} But, more important: my regimentation has two implications which amount to strengthening the conception of ontological emergence:\\

(1)~~{\it The requirement that $T_{\sm t}$, and a sufficiently good interpretation of it, must exist.} My conception makes emergence less pervasive than it might at first sight appear, because of the requirement that $T_{\sm t}$ must be a {\it bare theory}, presented in the same form as $T_{\sm b}$: usually as a triple, subject to the requirements in Section \ref{ern}. Thus one cannot take an arbitrary bottom theory, apply to its elements ${\cal S}$, ${\cal Q}$, and ${\cal D}$ some arbitrary map one calls `link', and claim that one has emergence: for one is not guaranteed to get as the range of the map labelled `link' a {\it bare theory}! 
Likewise for the interpretation, which must be sufficiently good, in the sense of the preamble of Section \ref{ern}: it must be empirically adequate and the interpretation map must be surjective. Again, some arbitrary map one calls `link' will in general not secure empirical adequacy of $T_{\sm t}:=\mbox{ran}(\mbox{link})$, nor will the thus-obtained theory describe the whole domain. 

(2)~~{\it The requirement of novel reference, for ontological emergence.} Interpretations that are genuinely novel (because they refer to novelty in the physical system) and, at the same time, empirically adequate to within required accuracies, need not be abundant. The reason is as follows:

Ontological emergence requires that we have a case of idealisation (see Section \ref{nnr}), so that there is a physical system to which the map refers. And this is, in fact, a considerable restriction on the linkage map: Norton (2016:~p.~46) states that `merely declaring something an idealization produced by taking a limit is no guarantee that the result is well-behaved.' In other words, `the limit operations generate limit properties only. They do not generate a single process that carries these properties, for these properties are mutually incompatible' (p.~44).\footnote{Norton (2016:~Abstract, and p.~46) has for example argued that there is no limit process that can be identified as a thermodynamically reversible process (as is standardly assumed by the notion of thermal equilibrium). According to Norton, the notion of `equilibrium', as construed in thermodynamics, ascribes contradictory properties to the system. This amounts to denying that thermodynamically reversible processes are idealised processes at all. On this reading, thermodynamics (as standardly construed) could not emerge from statistical mechanics, because no actual physical system (real or fictitious) can be in thermal equilibrium, on the standard construal of the notion. Whatever one thinks of this particular example (for an alternative analysis, see Lavis (2017)): Norton's general argument indeed shows that my requirement, that there is ontological emergence only if the composition of the linkage map and the interpretation map, $i_{\sm t}\,\circ\,\mbox{link}$, maps to an idealisation, is a substantial requirement. Indeed the example shows one way in which the top theory's interpretation map can fail to refer: namely, if the linkage map is such that the domain of application of the top theory is ascribed contradictory properties.\label{Ncontra}}

Ontological emergence thus crucially depends on the choice of ranges: for $D_{\tn t}\nsubseteq D_{\tn b}$ is the mark of novelty of reference. The ranges are to be determined by the theory's best interpretation, which is subject to empirical adequacy and the other constraints mentioned in Section \ref{ern} (see also below).\\

To what extent do we get a clear-cut criterion of ontological emergence, as I promised in the Introduction? While I do not claim that formulating ontological emergence formally {\it automatically} dispels problems of ontology, which can be subtle (see Section \ref{mando}): the criterion does give us a recipe for assessing whether ontological emergence obtains: it requires us to formulate theories, their interpretations, and the linkage relations, as formally\footnote{Crucially, `formally' is here taken in the official sense given in the Introduction, as in `mathematical philosophy': {\it not} in the sense of `uninterpreted'. } as possible. Once that work is properly done, the decision about ontological emergence follows as the non-meshing of linkage with interpretation, provided both theories refer. The challenge lies in adequately and formally formulating interpretations and linkage maps. To illustrate, let us return to the example of water. Consider the questions: 
 
Is liquid water ``more than'' the sum of molecules and their interactions? Should we consider liquid water as different from its molecular basis? What distinguishes the relation between the liquid state and its molecular basis from its cousin state, ice, and {\it its} molecular basis? 

My proposal does not do away with the need to ask such questions, at one point or another. These questions will get implicitly answered once sufficiently precise interpretations have been developed.
But, on my account, these are not the important questions for the philosophical account of emergence. The recipe is to develop as accurate as possible interpretations of the theories, subject to empirical adequacy and the other requirements from Section \ref{ern}, and to then allow those interpretations to answer the question for us. 

Interpretations allow for comparisons: one interpretation is better or worse than another because it covers more or fewer cases, and is more or less empirically adequate and precise, etc. So, one now asks: is there, by the interpreted bottom theory's own lights, such a thing as `liquid water'? Or does the range, $i_{\sm b}\,(T_{\sm b})$, only contain items such as `the collection of interacting molecules'? Here, one should not try to judge, {\it independently of interpreting theories,} whether `liquid water' is the same as, or can be empirically distinguished from, `a collection of molecules'. The right question is whether the interpreted bottom theory, in its range, $i_{\sm b}\,(T_{\sm b})$, has the hydrodynamic substance `liquid water' as an element---for recall, from Section \ref{nornc} (1), that we could have epistemic emergence if the range of the top theory was a subset of the range of the bottom theory. Now if our bottom theory is a theory of molecules and their interactions, it does {\it not} have `liquid water' in its range: and so water is, indeed, ontologically emergent in the top theory, $T_{\sm t}$, with respect to its molecular basis in $T_{\sm b}$. For on an interpretation containing only collections of molecules, there is never a continuous state of matter, however numerous the number of molecules, and however small the size of the molecules, may be. This is because a discrete set of molecules and a continuous medium are {\it referentially distinct} (recall that interpretation makes reference to concrete objects: it is not just more theory!). Such an interpretation may not need to distinguish, for its limits of accuracy, between $10^{40}$ and $10^{40}+1$ molecules (the number of molecules may only be defined up to $5\%$ accuracy, say). But it {\it will} distinguish a discrete from a continuous medium, because these are different kinds of objects, and so they constitute different domains.

The above is best understood when formulated using intensional semantics (see Section \ref{ern}). The hydrodynamic and molecular theories have the same extension but different intensions (liquid water vs.~a large collection of interacting molecules). The intensions differ because the theories are different and, although they may share many common terms (such as `volume', `pressure', `temperature', etc.), `liquid water' is not one of the terms that they share. Nevertheless, they both refer to the same experimental phenomenon in our world and, within certain margins of accuracy, their predictions are the same.

Many examples of emergence are of this type: the theories have the same extension, but different intensions.
Of course, it is also possible that both the extensions and the intensions differ.\footnote{This possibility gives rise to the various cases considered in Section \ref{whichs}. Theories with the same extension but different intensions were previously considered, in the debates between the Nagelian vs.~Kuhn-Feyerabend positions on reduction, by Scheffler (1967:~pp.~60-63) and Martin (1971:~pp.~19-21). Nagel (1979:~pp.~95-113), together with his (1961:~pp.~342, 366-374) can be taken to hint at the relevance of this distinction for emergence. However, I {\it disagree} with these authors in several other aspects: not least, Scheffler's (1967:~p.~57) statement that `for the purposes of... science, it is sameness of reference [extension] that is of interest rather than synonymy [intension]'. \label{int-lit}}

The distinction between extensions and intensions also further clarifies the criteria of identity of domains, from Section \ref{ern}. For sameness of extension is chiefly (of course not solely!) determined by an empirical procedure, while sameness of intension is conceptual, and determined from within the theory\footnote{See for example Martin (1971:~p.~21).} (again, not without empirical input!).

Finally, I stress that there can be no ``fiddling'' with the interpretation in establishing whether there is emergence. This is the strategy I will adopt in my case studies in Sections \ref{imct}-\ref{casemm}: interpretations must be fixed (by other criteria) before one asks about emergence. This is especially true for {\it intensions.} In other words, ontological emergence can only be predicated relative to interpretations that are sufficiently good (on the criteria given in Section \ref{ern}): and different interpreted theories must be assessed against each other employing the usual evaluation criteria given by the philosophy of science and in scientific practice. Thus my proposal is to decide for one's best interpretation once and for all using independent methods, to take the interpretation literally, and to stick to it when enquiring about emergence. Changing one's interpretation in the course of assessing emergence is liable to lead only to confusion.

{\it Reduction and emergence.} One should not confuse the question of emergence we are asking here with the question of reduction. We already agreed that, in this example, there {\it is} reduction between the bare theories: emergence and reduction are compatible! (see Section \ref{nornc}).\footnote{Butterfield (2011:~\S3.1.1-(4)) reminds us that, in cases of reduction (via definitional extension) of $T_{\sm t}$ to $T_{\sm b}$, $T_{\sm t}$ does not extend the domain of quantification of $T_{\sm b}$ (``it has no new objects''). I of course accept, as everyone must, this verdict: but it does not bear on interpretation maps, which in putative cases of emergence are {\it not} related by reduction. 
The question of emergence is whether the pictures of the world, that we get from the two interpretations, mesh with the reduction relation between the bare theories.} The point is precisely that reduction, when it obtains, is a relation between the {\it bare theories},\footnote{As Nickles (1973:~pp.~183, 185, 193-194) emphasises, ontological issues are not central in the {\it reductions} done by physicists, and most cases they consider are not cases of ontological reduction. Rather, the issues are typically explanatory, heuristic, and justificatory. My weak notion of reduction agrees with this.} while ontological emergence is in the range of the {\it interpretation}. Thus, {\it the reduction of the bare theory cannot imply a putative reduction of the interpretation.}\footnote{Recall, from footnote \ref{ontred}, the distinction between reduction of bare theories (i.e.~formal reduction regardless of interpretation, which is my default meaning of `reduction') and the more controversial notion of ``ontological reduction''. The latter is controversial because, in all the cases of emergence here considered, the intensions differ. That the intensions differ is a point about which Feyerabend (1963:~pp.~3-39) and Nagel (1979:~p.~95-113) agree, even though they draw different conclusions. See, for example, Schaffner's (2012:~p.~548) review, where he explicitly states that the bridge principles (which he calls `connectivity conditions') are {\it extensional} not intensional, since the meanings differ.\label{ontred2}}

\section{Further Development}\label{fdev}

In this Section, I take up three further questions that my framework for emergence prompts. In Section \ref{whichs}, I give further details about the systems that are related by emergence. In Section \ref{mando}, I explicate the sense in which the kind of emergence described here is ontological. In Section \ref{comparrw}, I compare to other recent work on emergence.

\subsection{Which systems?}\label{whichs}

In this Section, I will further specify the systems that are related by the emergence relation. This question is prompted by the condition (ii), in Section \ref{emred}, that the linkage map relates bare theories for which appropriate interpretation maps exist, which map to physical situations or systems. 

To answer this question, let us recall the following distinction, from Norton (2012:~p.~209) at the beginning of Section \ref{nnr}: The {\it target system} is the system we aim to describe: the chunk of material in the lab, the gas of Hydrogen ions (composite particles!), which are fed to the LHC through its linear accelerator, etc. The {\it idealising system} is the system to which our idealising theories refer, and it may well be distinct from the target system. 

The question I will ask is about the comparison between the idealising system(s) and the target system. I will address this question by looking at the empirical adequacy of the idealising sytem(s) relative to our target system, i.e.~by asking how accurate, or exact, is their description of the target system.\footnote{By `description', I do not here mean to talk about the `truth' of the idealising theories in a deep sense nor, for the purposes of this Section, about their reference.} Thus the question is simple: how empirically adequate, or exact, is the description of the target system given by the idealisation?

Suppose that we have a theory, $T_{\sm b}$, describing a certain target system within given limits of accuracy. Then the domain of the theory, $D_{\sm b}$, contains a system that is close to our target system. Typically, the system described by the domain of the theory will not be the target system itself, but a closely related system---an idealised system. Thus the domain $D_{\sm b}$ does not literally describe our world, at least not in full detail, but a physically possible world---it gives an inexact description of the target system (as both idealisations and non-idealising approximations do). In some cases, however, we may not be able to tell the difference, and we may just identify the target system with the idealising system---but this is not a generic situation, since future experiments may expose the inexactness of the description. The same can be said of $T_{\sm t}$ and {\it its} domain, $D_{\sm t}$: it gives an inexact description of the {\it same} target system, even though it refers to an idealising system that is, in principle, distinct from it. 

As I said above, I will not here attempt to give an account of under what conditions the idealising system can be identified with the target system. Rather, my aim is to assess how accurate the descriptions are that are given by the bottom and top theories, i.e.~about their empirical adequacy. So, we have three possible situations: \\
\\
(1)~~The bottom theory gives the more accurate description of the target system.\\
(2)~~The top theory gives the more accurate description of the target system.\\
(3)~~The two theories are equally empirically adequate. 

It is worth noting that, while (1) is often taken to be the default option, this need not be so. As we will see in the example of Section \ref{mmpp}, if we are interested in describing massless particles, then $T_{\sm t}$ is the more accurate theory, and the role of $T_{\sm b}$ is largely explanatory and unifying, thus a case of (2). Which of (1)-(3) is the more appropriate option is of course an empirical question.

This discussion sheds light on the nature of ontological emergence: ontological emergence is simply a matter of {\it comparing physical situations or systems}, whether real or fictitious, at our world or at another physically possible world, characterising the same target situation or system (although we may compare the target system under different conditions, such as different values of the temperature: see the example in Section \ref{imct}). 

\subsection{Ontology and metaphysics}\label{mando}

In this Section, I want to answer the following question: In what sense is ontological emergence {\it ontological}? 

As I said in the Introduction, I construe `ontology' here in the straightforward sense of `the ontology of a scientific theory': which, crucially, I take to be more than mere semantics, since the theories in question are subject to the requirement of empirical adequacy: so in particular, they must describe the same target system (cf.~Section \ref{whichs}). We are interested in physical, and not merely mathematical, possibilities. Thus I construe ontology, in a somewhat limited sense, as the subfield of metaphysics that is concerned with what is: but not in Quine's narrow sense of `to be' as in `to be the value of a bound variable'. My sense of `ontology', on which I will expand further below, comprises both what Quine (1951:~pp.~14-15) dubs the `ontology' of a theory (the domain of objects that the theory quantifies over) and its `ideology' (i.e.~roughly: the meaning of the theory's non-logical vocabulary, here understood as elements and relations in the domain of application, which are part of the world).

A first comment to make is that the worlds that I have been describing are {\it not} to be (naively, and wrongly!) identified with {\it the world as it is in itself}---whatever that Kantian phrase might be taken to mean. 

Second, there is an interesting metaphysical project of (a) studying {\it how} the entities postulated by scientific theories can be, even if (b) we have not yet decided whether and how they {\it exist}, i.e.~even if we have not yet decided how the idealised systems characterise the target system. That is, we can assess the question of the being and properties of those entities as idealisations, according to the theory, before we ask about their existence at our actual world.\footnote{As noted in De Haro and Regt (2018:~\S1.1), the conception of interpretation used here is {\it not} model-theoretic, although it does give a model-theoretic notion for restricted classes of interpretations. One salient difference is that the current notion allows for non-linguistic interpretations of the terms of the theory (e.g.~by means of Feynman diagrams, or interpretation by ostension).} The latter question, (b), of course requires commitment to a specific metaphysical position: the realist will say that those entities exist (approximately) as postulated by the theory, while the non-realist may say that these entities are fictions, or reduce to combinations of human perceptions. Also, realists may disagree amongst themselves about a metaphysical construal of those entities: think, for example, of Quine's (1960:~\S12) {\it referential indeterminacy}, according to which the linguist, upon hearing the native utter the word `gavagai' while pointing at a rabbit, might translate it as `rabbit'---while still being at a loss whether to construe the objects as rabbits, or stages, i.e.~brief temporal parts of rabbits, or mereological fusions of spatial parts of rabbits. My construal of `ontology' is limited in this sense, that `rabbit' refers to the same element in the domain of application as `gavagai', i.e.~they refer to the same object even before we ask what further metaphysical categories the native appeals to in their understanding of `gavagai'.

Another example of (b) is as follows: the quidditist will construe the novelty of the domains as a difference in the domains' {\it properties}: and the comparison will be based on the possibility of the primitive identity of fundamental qualities (for example, natural properties: cf.~Lewis (1983:~pp.~355-358)), across domains or worlds---irrespective of how the quidditist construes individuals in his or her ontology. On the other hand, the haecceitist will construe ontological novelty as a difference in the domains' objects, or individuals, irrespectively of how he or she construes properties in their ontology.\footnote{See Black (2000:~p.~92).}

While these differences could, to a large extent, be already built into the interpretations, I have chosen to keep things general and not give a specific metaphysical construal of the domains (other than saying that they are sets with elements and relations, and that they can be intensions or extensions). 

Question (b), applied to our case of distinct domains of application, $D_{\sm t}\not=D_{\sm b}$, means that different metaphysicians can agree about the parts of $D_{\sm t}$ that are not in $D_{\sm b}$, according to Eq.~\eq{dnotd}, since these are structured sets. But different metaphysicians will construe the differences differently---over and above the comparison given by the description of the top and bottom domains as structured sets.

But the ontological project in this paper does not require that we answer question (b): for the realist and the non-realist alike must agree in constructing the ontology of the theory, before this ontology can be, for example, declared real or reduced to perceptions. For example, realists and constructive empiricists can agree about the interpretation of a theory, even if they have different degrees of belief in the entities that it postulates (van Fraassen (1980: pp.~14, 43, 57)). The idea here is that working out the ontologies of scientific theories, the way they are interconnected, and their logical structure, is a different project from explicating how the elements of that ontology actually exist.\footnote{An anonymous referee objected that the distinction between epistemic and ontological emergence is unimportant for a theory of emergence, because it depends on whether one adopts a realist or, for example, an empiricist or a conventionalist framework. However, I am arguing that the distinction {\it is} important, even before we address the debate between realism and its rivals. For even the empiricist has to admit that any interpreted scientific theory assumes certain ontological facts about the world: even if a world only of phenomena, to which the empiricist is minimally ontologically committed. For example, the empiricist assumption that all that our theories describe are regularities rests on some ontological assumptions: viz.~that there are regularities in the world for our scientific theories to describe, and that those regularities do {\it not} point to any deeper ontological structure. Furthermore, Ladyman has argued that van Fraassen's belief in the empirical adequacy, rather than the truth, of theories requires an objective modal distinction between the observable and the unobservable: so that the empiricist cannot avoid modal talk. For a summary of that debate, see Ladyman and Ross (2007:~pp.~99-100). \label{empiricism}} 

My position here bears some resemblance with what A.~Fine (1984:~p.~96-98) has called the `core position' that realists and non-realists share: both accept the results of scientific inverstigations as, in some sense, `true', even if they give a different analysis of the notion of truth. It corresponds to the contrast that Schaffer (2009:~p.~352), Corkum (2008:~p.~76) and K.~Fine (2012:~pp.~40-41) have made, in the context of scientific theories, between a neo-Aristotelian metaphysical project (of enquiry into {\it how things are}) vs.~a Quinean project of strict enquiry into {\it what exists} (in Quine's very narrow sense as domain of quantification), according to our best theory, appropriately regimented.\footnote{I endorse this contrast, even if it may not be entirely faithful to the historical Quine and, especially, Aristotle (hence the word `neo-Aristotelian').} The former project permissively allows for things, and categories, to appear in our ontology, that we might one day come to reject as literal parts of our world. Those things are, in some sense: even if they do not exist in the literal sense in which the theory says they do (for example, for the reasons given in Section \ref{whichs}: the systems described by the theory are idealisations of the target system). Thus my position, which is based on a limited but straightforward reading of the ontology of a scientific theory, is closer to the `jungle landscapes and coral reefs' (Richardson (2007)) of the neo-Aristotelian project than to the desert ecosystems that Quine's advocacy of first-order logic suggests.

\subsection{Comparing to other recent work}\label{comparrw}

Novelty of reference, and the distinction between intensions and extensions, are of course not new notions. But, so far as I am aware, they have not been used before to define a conception of ontological emergence.\footnote{For some early ``hints'', see footnote \ref{int-lit}. However, my account differs significantly from the language-based logical empiricist ideas.} In order to (a)~justify this statement, (b)~compare my framework to other approaches in the literature, and (c)~explain how the framework improves on existing accounts, I discuss, in this Section, a number of recent influential accounts of emergence. I first compare my account with some recent general works on emergence (Section \ref{ppw}) and then with more specific ones (Section \ref{swe}).

\subsubsection{Some work on emergence in general}\label{ppw}

Humphreys' (2016) account of emergence distinguishes ontological, conceptual, and inferential emergence: but in none of these conceptions is novelty explicitly construed as novel reference. Rather, Humphreys' ontological novelty seems to be entirely theory-free, as well as realistic (`Ontological approaches consider emergent phenomena to be objective features of the world':~p.~38). This seems, in part, related to the centrality of causation in Humphreys' account of emergence. Indeed, in Humphreys (1997:~p.~8) he defined the emergence of properties in terms of their causal powers.\footnote{A question that arises if one considers emergence in terms of causal powers is the problem of top-down causation (see Ellis (2016)). Humphreys (2016:~p.~70) solves it by introducing the notion of (non-mereological) fusion. But I will not need to discuss this here, since my account does not involve causation.} My conception of ontological emergence is objective but at the same time theory-relative, and its relation to the actual world is not a one-to-one correspondence relation, and it does not require scientific realism, as I have argued in the previous Section. 

Nevertheless, ontological emergence as here defined {\it is} related to Humphreys' ontological emergence, which comprises property emergence, object emergence, and nomological emergence (i.e.~the fact that there are autonomous laws that govern the top domain). The domains, as defined here, will typically contain properties and objects (see Section \ref{mando}).

Humphreys (2016:~p.~40) says that an entity is conceptually emergent, with respect to a theoretical framework, iff a conceptual or descriptive apparatus must be developed in order to effectively represent that entity. This corresponds, roughly, to my epistemic emergence: since the emergence is not in the world, but in the bare theory and its interpretation map. \\

Other accounts in the philosophy of physics tend to either ignore the distinction between epistemic and ontological emergence, or to even give incorrect accounts of it. 

I discuss here one important recent account, with which I will disagree, namely Crowther (2016:~p.~40).

Generally, I sympathize with Crowther's (2016:~p.~42) notion of emergence, which agrees with the philosophy of physics consensus about emergence consisting of two main factors: namely, Section \ref{emred}'s `delicate balance' between dependence and autonomy.

However, Crowther---perhaps motivated by accounts of emergence that focus on the `lack of derivability'---claims that the main factor distinguishing ontological and epistemic emergence lies in the ideas of `deduction and derivability': `the tendency to think of emergence as a failure of reduction (or derivation) is related to the desire to classify cases of emergence as either ontological or epistemological' (p.~40). More explicitly, on p.~42 she {\it defines} ontological emergence as `the thought that this failure of reduction (in whatever sense is meant) is a failure {\it in principle}... This stands in contrast to {\it epistemological emergence}, which is a failure of reduction {\it in practice}.' 

Crowther's notions of ontological and epistemic emergence are, I submit, incorrect; and so her reasons for dismissing that contrast are ill-motivated. First, not every failure of reduction leads to ontological emergence---nor is failure of derivation indeed required for ontological emergence, as I already discussed in my criticism of Broad in the Introduction. For, as I have argued,\footnote{See the Introduction, the end of Section \ref{nornc}, and Section \ref{oee}.} the question of reduction between bare theories does not bear on the epistemic vs.~ontological distinction: since we can have ontological emergence with a linkage map that is reductive. And second---even admitting that the jargon in the literature is varied---the `in principle' vs.~`in practice' distinction is not distinctive of the ontological vs.~epistemic contrast, but is a difference that appears within a given type of emergence, and it is usually called `weak vs.~strong emergence': cf.~for example Guay and Sartenaer (2016:~pp.~299-300). Thus we can have weak and strong emergence of both epistemic and ontological types: and we can have failure of reduction in practice or failure of reduction in principle in both cases of ontological and of epistemic emergence. 

The discussion in Section \ref{emred} argued that `novelty', rather than `irreducibility', is a better and more general starting point for a characterisation of emergence. Section \ref{nnr} then specified ontological emergence as novelty of reference, thus clarifying how to get from a generic conception of emergence to ontological emergence.

This tendency to ignore the epistemic vs.~ontological distinction, or to conflate one with the other, runs through more of the recent literature. So, for example, Butterfield (2011a), when describing emergence in an example of limiting probabilities, slips in the word `surprising' into the definition: `For me, [emergent behaviour] means behaviour that is novel or surprising, and robust, relative to a comparison class.' While novelty here can of course be of any kind, `surprising' refers to our epistemic situation: and so, this definition is ambiguous as to whether it construes emergence as ontological or as epistemic. In the same vein, Humphreys (2016:~p.~32) writes that `Novelty is usually required to be of a striking kind---not just any novelty will do'. Humphreys seems to be critically quoting a consensus here, rather than endorsing it, for he adds: `since what counts as an acceptable kind of novelty is rarely, if ever, specified, it is difficult to capture that additional aspect with any generality'. I endorse this criticism of the literature. Indeed I do not see why being `surprising' or `striking', or some other such qualification---even if perhaps indicative of scientific salience---is {\it necessary} for emergence: since it brings in an unwarranted subjective element.

\subsubsection{Some work on specific emergence: the case of phonons}\label{swe}

I will now compare to a more specific example of emergence, in Franklin and Knox (2018), which would appear to contravene some of my conclusions. This will also be an interesting test of my framework, which is able to give various verdicts, to which the examples to be worked out in Sections \ref{imct}, \ref{mmpp}, and \ref{casemm} will further add.

The question at issue, comparing with Franklin and Knox (2018), is whether `objects can emerge without the existence of an ideal system in the top theory's interpretation'.\footnote{I thank an anonymous referee for suggesting to engage with this apparent conflict. A similar comparison could be made with Wallace (2010).} My account clearly says: No, not if this is to be a case of {\it ontological} emergence. For recall, from Section \ref{nnr}, that the linkage map must have an idealisation in its range.

But, as I will next argue, I am nevertheless not in conflict with these two papers. Rather, apart from peripheral disagreements, we simply adopt {\it different uses} of the terms `idealization' and `approximation', as I will explain below, after I introduce this work.

\begin{figure}
\begin{center}
\includegraphics[height=2cm]{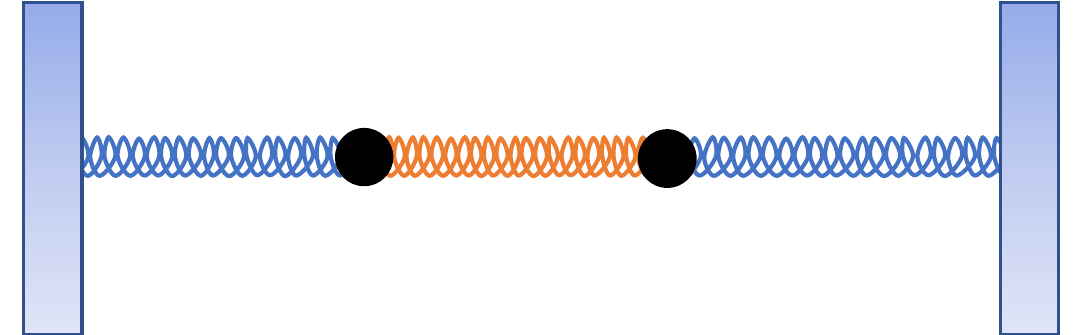}
\caption{\small Three coupled springs with two equal masses, fixed to two outer walls. The outer springs have the same spring constant, different from that of the middle spring.}
\label{springs}
\end{center}
\end{figure}

Franklin and Knox (2018) discuss the emergence of phonons, i.e.~collective harmonic excitations of large chunks material in a crystal, by an appropriate change of variables, combined with a suitable approximation, from the underlying atomic description. They introduce their case study with the analogy of two particles of equal mass coupled through three springs. Here, two of the springs are fixed to outer walls, and the middle spring connects the two masses that oscillate in between; the outer springs have the same spring constant, which differs from the middle spring's constant (see Figure \ref{springs}). Using Newton's equations including the Hooke term, one gets a system of two coupled second order linear differential equations in the displacements, i.e.~one equation for each mass. Now although the two equations are coupled, they can be uncoupled by going to `normal mode' coordinates, which are linearly independent combinations of the two displacements. These uncoupled equations describe two simple harmonic oscillators with different frequencies, whose solutions are straightforwardly given by harmonic oscillations in time, viz.~sines and cosines. 

There are two aspects to this being a case of emergence: Franklin and Knox argue that the harmonic behaviour is novel, and could not have been found without the change of coordinates. They further imagine making the central spring out of material that glows when compressed: one then seeks to explain the frequency with which the spring lights up from the dynamics of the springs. They argue that, in the original displacement variables, any explanation of the phenomenon will require two variables and two equations, while in the normal mode variables there is a {\it single} equation associated with that frequency. Thus the normal mode description `abstracts away from irrelevant details in a way that an explanation on displacement variables cannot' (p.~70). The appeal here is to explanatory power. The normal mode explanation is {\it more explanatory} because it focusses on the relevant physics. The second aspect to the example is that anharmonic corrections can be included. If the springs are not ideal, Hooke's law is corrected by higher-order (anharmonic) terms or even velocity-dependent terms. One can deal with this in either set of variables: once one has achieved the change of variables to normal modes, the corrections can be written down in terms of the normal modes.\footnote{Most of the computational simplicity is then lost, but conceptually the normal mode description is still clearer, if the corrections are small.}

The phonon description is a generalisation of this: here we view the atoms of a crystal as located on the sites of a three-dimensional lattice, and connected by springs in various directions. It seems that the differences with the previous model are mainly technical rather than conceptual. Now there are a high number of particles, which can move in all directions, and to reach the decoupling of the equations, i.e.~find the normal modes, the trick is to do a discrete Fourier transformation to momentum space. The equations that one then gets for each of the normal mode variables are ordinary uncoupled harmonic oscillators. The harmonic vibration modes of these oscillators are called `phonons'. They are thus called because they `naturally describe the crystal not in terms of individual atomic displacements, but in terms of aggregate properties of the whole supercell' (p.~70). They describe the vibrations of the crystal itself. 

Overall, Franklin and Knox seem to make three main claims about phonons:

(1)~~The change of variables is {\it computationally} powerful. 

(2)~~The change of variables is {\it explanatory} powerful.

(3)~~There is an {\it approximation} involved, and no `essential idealisation', i.e.~no idealisation that cannot be de-idealised without losing the phenomenon in question. In plain language: the phonons are still there in the crystal even in the presence of small corrections. These corrections are larger when the temperature increases, because the atoms then tend to fluctuate more.

I take it that the claim of emergence follows mainly from (2) and (3). (2) is the novelty condition, here construed as `novel explanation'. Like in the case of the simple system of coupled springs, the normal modes give a novel explanation of the behaviour. And (3) means that the emergent behaviour is robust, i.e.~not washed out by small anharmonic terms due to the increase of the temperature.

As I will explain in a moment, I agree with all of the above, which does {\it not} contradict my own account of emergence.

Early in the paper (in Section 3.1, pp.~72-73), Franklin and Knox discuss whether phonons are emergent. They have two main arguments for this: First there is the argument that phonons `are really out there' (p.~77). Second, there is a formal analogy with the emergence of particles in quantum field theory. For once one Fourier transforms the phonons and quantises them, there is a strong formal resemblance with Fock space quantisation in quantum field theory. 

Their Section 3.1 is---in its wording---certainly suggestive of the idea of ontological emergence: however, Franklin and Knox never use this exact phrase---to their credit! Thus there are two interpretations of the Section's analysis, of which the first is a weaker claim than the second:

(i)~~There is {\it emergence,} but this emergence does {\it not} need to be qualified as `ontological'. There is merely the claim that `phonons should be taken as {\it ontologically serious} as quantum particles' (p.~72, my emphasis). In other words, ``phonons are really {\it out there,} they are part of the theory's {\it ontology'}'. 

(ii)~~The emergence of phonons in the way described is {\it ontological emergence.} 

I believe that the literal reading (i) is the correct one: First, the authors never use the phrase `ontological emergence'. Second, their arguments only entail (i) not the stronger (ii). They are arguments for the {\it existence} of phonons not their `ontological emergence' (a phrase which the authors neither use nor define). In a moment, I will argue against the truth of (ii). But first, let me explain (i) a bit more. 

I take it that Franklin and Knox are discussing {\it epistemic} emergence: it is the emergence of a description that is very useful because it is both powerful and explanatorily more parsimonious (cf.~(1)-(2) above), since it focusses on the degree of freedom relevant to answer the question at hand, namely: What is the frequency of oscillation of the normal modes? But this is separate from the question whether phonons are ``really out there in the world''. {\it Of course} phonons are out there as real patterns! Clouds, dunes, waves are all real patterns, even though they might not all be cases of ontological emergence. Phonons are just as real. But that is different from the question whether the pattern counts as {\it ontologically novel:} and relative to {\it what} it may count as novel. For example, there is a real difference between the behaviour of the lattice at high and at low temperature. At low temperature, the oscillations are harmonic and are described by phonons, while as we increase the temperature the anharmonic terms start to become important. In other words, phonons are ontologically emergent with respect to one particular bottom theory (the high-temperature crystal as we lower the temperature), but only epistemically emergent with respect to another bottom theory (the atomic theory, where we keep the temperature fixed but compare different scales)---as I will now substantiate. 

Thus I now argue against the truth of (ii). I will argue that their analysis is not an analysis of ontological emergence (which I think would be reading notions into the paper that it does not mention). 

The general account of `novelty' that Franklin and Knox defend is `explanatory novelty',\footnote{Knox (2016) views explanation in terms of abstraction, which she argues is compatible with standard accounts of explanation. She points out that certain choices of descriptive quantities (especially, choosing the right variables) can lead to novel explanations that are not {\it merely} abstractions of some more detailed picture.\label{Knoxexp}} but `explanation' is ambiguous between being an epistemic and an ontological notion, and so it does not by itself secure that emergence is ontological.\footnote{Knox (2016:~p.~58) points out this ambiguity when she writes that her account of explanation is in some ways weaker than standard epistemic accounts of emergence, and in some other ways stronger than an epistemic account. I agree that explanations can latch on to ontological facts; but, as I argued above, in order for explanation to lead to an account of ontological {\it emergence,} close attention must be paid to the relation between the top and bottom theories.\label{ontexp}} 
I concur with Knox (2016) that explanation is important for emergence; but I see explanation as the {\it aim} of judgments of emergence: declaring that a phenomenon is emergent adds to the phenonemon's interest because it explains something about it: namely, it points to the phenomenon's rootedness in the bottom theory, while securing its independence. But explanation is not the defining mark of emergence: on pain of trivialising the important question {\it Why is emergence explanatory?,} i.e.~making it a matter of definition! 

Explanation, like novelty, can be either epistemic or ontological: if I explain the conservation of the angular momentum of a rotating body using polar coordinates, I will have reached a simple explanation that focusses on the relevant variables. But I can explain the same facts using Cartesian coordinates. My choice of coordinates does not change the body's ontology: for it will spin in the same direction and at the same rate, regardless of this choice. Explanations of this type are matters of our description: it should not make any difference, for this local problem, whether I choose one set of coordinates or another. The theory, and the motions that result from solving the equations, are the same. 

Applied to the change of coordinates involved in the case of the coupled springs and in the case of the crystal, it is no different. The simpler explanation in terms of normal modes does not capture more facts about the domain of application than does the explanation in terms of displacement variables: it captures the same facts. There is only one theory, formulated in different variables. 

I think there are two very clear ways to see this. The first is Franklin and Knox's thought experiment, mentioned earlier, of making the middle spring glow when it contracts with a certain frequency. It is a fact that the spring will glow when it compresses with its normal frequency, regardless which coordinates I choose to predict this fact. The spring and the glowing mechanism know nothing about my description in coordinates. Thus the property `glows' does {\it not} emerge ontologically when we change coordinates: it glows {\it independently} of my description, and if a description is empirically adequate it must give an explanation for the glowing, i.e.~it had better predict the value of the frequency at which it glows. I think it would amount to ``coordinate fetishism'' to claim, in such local cases,\footnote{Notice that we are here discussing a simple question about a linear chain of springs, and not a cosmological question about the state of motion of the whole universe.} that one set of variables `captures more facts about the world' than the other---and, as I already explained, I do not think this is what Franklin and Knox mean. 

The second way to see this independence of the description is as follows. The normal modes method is not the only way to get the frequencies of the normal mode oscillations. One can of course take recourse to linear algebra, and calculate the eigenvalues of the relevant $2\times2$ matrix associated with the set of equations. 
The eigenvalues are independent of the system of coordinates chosen, and in several important ways one might wish to say that calculating a determinant is simpler than looking for a suitable change of variables---it is also more powerful, since it can solve a larger class of problems more efficiently, and in a unified formalism. It not only calculates the frequencies---it gives the full solution, through a mathematical theorem that gives the solution in any desired, linear coordinate system. The explanation counts as more abstract, because it requires less details. 
Have I now reached an explanation of the frequencies that `captures more facts' about the harmonics? The answer is no: it captures the very same facts.\footnote{It does capture some more facts about the symmetries of the system; but this is besides the point.} This coordinate-independence of the matrix method also means that there is no such thing, in this case, as ``the right variables'' for solving the problem---and on this point I disagree with Franklin and Knox.

So far I have argued that (i), rather than (ii), above is the correct interpretation of Franklin and Knox (2018), i.e.~that the emergence they discuss is epistemic not ontological. 

Let me now comment on the robustness of emergence, i.e.~point (3) above: specifically, on the role of `idealisations' and `approximations' in emergence. As I said before, I think that any putative substantive disagreement between me and Franklin and Knox is mainly verbal, thus apparent. For they seem to interpret Norton's phrase `idealisation' as `essential idealisation', i.e.~one that cannot be de-idealised without losing the phenomena in question. This leads them to conclude that their phonons are not cases of idealisation, but rather non-idealising approximations, since there is no essential idealisation involved---the phonons are still there if we increase the temperature a bit. But it seems to me that this is a misreading of Norton's intended meaning of `idealisation', which nowhere includes the condition that it should be `essential'---and I do not think that any of his examples, like the one of the caloric theory of heat that I quoted in Section \ref{nnr}, point in the direction of essential idealisation. I take it that the confusion is caused by Norton's somewhat unusual use of `approximation', cf.~footnote \ref{nia}. I, at any rate, interpret him as merely saying that, in cases of idealisation, the idealisation {\it may} still be there approximately (my meaning of `approximation' not his!) once we go away from the limit. The idealised system, if robust, will still give a good description when we include corrections. 

This is tied up with the idea, which I take from both Norton and Butterfield, that there is already emergence ``before the limit''. The way this plays out in my examples of Section \ref{mmpp} and \ref{casemm} is that the limits that I will take are all smooth. Thus the ideal system is already visible before we get to the limit, which is precisely the meaning of emergence ``before the limit'' in Butterfield (2011a:~pp.~1069-1073). It is {\it not} the case, in those two examples, that the ideal system ``suddenly appears'' because of some singularity at the limit, and disappears if we go away from it---rather, it usually gets {\it deformed,} away from the limit. This means both that, on my notion of ontological emergence, the limits {\it need not be singular}, and that the idealisation {\it need not be essential}. (In particular, emergence with ordinary approximations, as in (iii) in Section \ref{emred}, is possible). This brings me much closer to Franklin and Knox's position than our use of words suggests. For this reason, I claim that the phonons of Franklin and Knox {\it are} idealisations, even though they are not essential idealisations.  Thus {\it approximation and idealisation are not incompatible.}

\subsection{How broad?}\label{broad}

In this Section, I will say a bit more about the intended scope of my framework for emergence. How general is it, and where does it apply? 

The intended scope should be clear from the title: I am dealing with the physical sciences. But let us look, beyond this general statement, more specifically at what the framework assumes. On the one hand, as I announced in the introduction, it assumes theories that can afford the notions of sets and functions, i.e.~formal structures whose interpretation can be modelled by maps of the kind discussed in Section \ref{ern}. Another key element that the theories discussed should afford is the notion of a linkage map, as discussed in Section \ref{emred}. As I remarked in footnote \ref{triple}, the notion of a theory presented as a triple is, on the other hand, {\it not} a strict requirement. Any other formulation 
will also do. However, some such general structural presentation of the theory {\it is} required.

These requirements can also be weakened in the following sense. As I mentioned in Sections \ref{emred} and \ref{onep}, emergence often takes place between the {\it models} of a theory (or models of different theories), rather than between theories. Many of the examples of emergence that physicists deal with concern not entire theories, but how specific solutions change as we vary a parameter, within the context of a single theory or class of related theories. We will encounter these two types of `model-emergence' (within a single theory and a class of related theories) in Sections \ref{imct} and \ref{mmpp}, respectively. 

But the disciplinary demarcation problem is of course unsolvable, and so the use of my framework is not so much defined in a specific discipline, but---as I mentioned above---by the possibility of using formal notions for given theories. I expect {\it both} that my framework, perhaps with appropriate case-specific adaptations, can be applied in some formal theories and models {\it outside} of the physical sciences (for example, molecular biology)\footnote{As I will mention in Section \ref{casemm}, random matrix models are used for many purposes outside the physical sciences.} {\it and} that it might {\it not} apply to some models, and perhaps theories, {\it within} the physical sciences that are not yet sufficiently formalised or developed. But no matter, say I: the model {\it will} apply in cases of theories that are sufficiently formal, regardless where one may find them. 

\section{First Case Study: Spontaneous Magnetisation in Ferromagnetism}\label{imct}

To illustrate the idea of emergence in physical terms, and to get some acquaintance with how referential novelty appears, I will here discuss a much-studied case: emergence in phase transitions, and in particular the emergence of spontaneous magnetization in ferromagnetism. My aim is not to give exhaustive details about the physics, about which there are very good reviews, but rather to illustrate the main ideas and the sense in which there is emergence.\footnote{For details about the physics, see for example example Binney et al.~(1992:~pp.~1-83), Chandler (1987:~pp.~129-131), Baxter (1982:~pp.~1-24, 88-126). For the effects of finite size scaling, see Cardy (1988:~pp.~1-7). For philosophical discussions, see: Batterman (2002:~Chapter 4), Humphreys (2016:~Section 7.1), Menon and Callender (2013) and Kadanoff (2013).} More detailed case studies follow in the two subsequent Sections.

\begin{figure}
\begin{center}
\includegraphics[height=4cm]{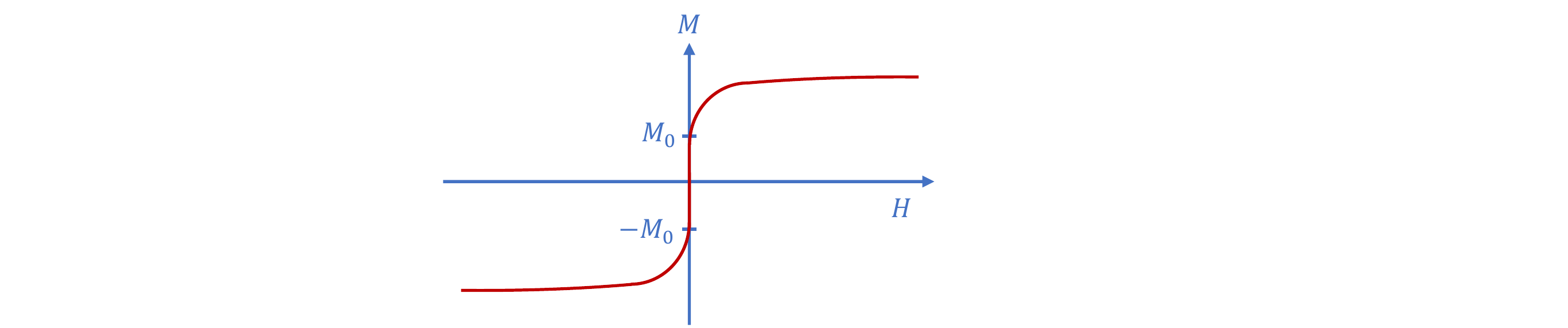}
\caption{\small Graph of $M(H,T)$ at constant $T<T_{\tn C}$.}
\label{magnetiz-a}
\end{center}
\end{figure}

Consider a bar of iron at low temperature, placed in a strong external magnetic field, $H$, that is aligned parallel to the bar's main axis. Now consider gradually decreasing the strength of the magnetic field until the field is completely absent. At zero field, the iron will be magnetized, i.e.~it will have a {\it spontaneous magnetization}, $M_0$, that is a function of the temperature. Now if we do the same but in the opposite direction, i.e.~if we reverse the sign of the magnetic field $H$, the magnetization $M$ will have the opposite sign, and so will the spontaneous magnetization, $M_0$. Thus the magnetization $M(H,T)$ must be an odd function of $H$. Its graph, for constant temperature $T$, is depicted in Figure \ref{magnetiz-a}. 
The iron can here be regarded as undergoing a phase transition at zero magnetic field, $H=0$, where the magnetization is discontinuous, and its sign changes from negative to positive.\footnote{This discontinuity is very much like the discontinuity in the density during a liquid-gas phase transition, as can be shown by calculating the van der Waals corrections to the ideal gas law: see Baxter (1982:~pp.~30-31). In an actual experiment, the discontinuity is smeared out, and the phenomenon of hysteresis occurs: namely, the magnetization, as a function of the magnetic field, is path- or memory-dependent. The existence of spontaneous magnetization below the critical temperature in real materials shows that we have here a robust physical phenomenon that does not hinge on theoretical idealisations.\label{actex}}

\begin{figure}
\begin{center}
\includegraphics[height=4cm]{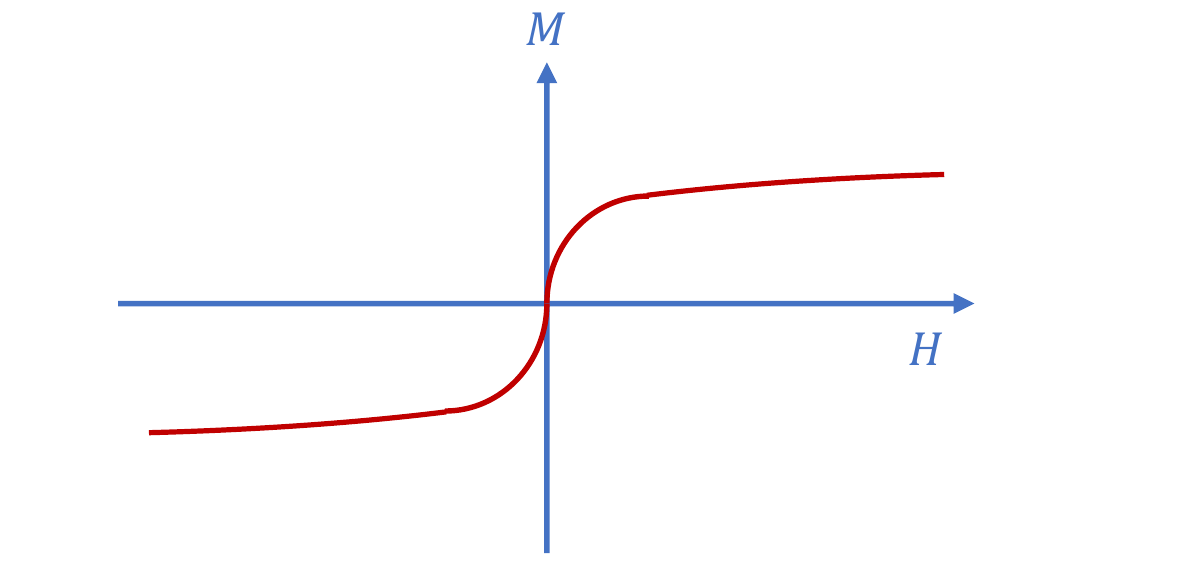}~~
\includegraphics[height=4cm]{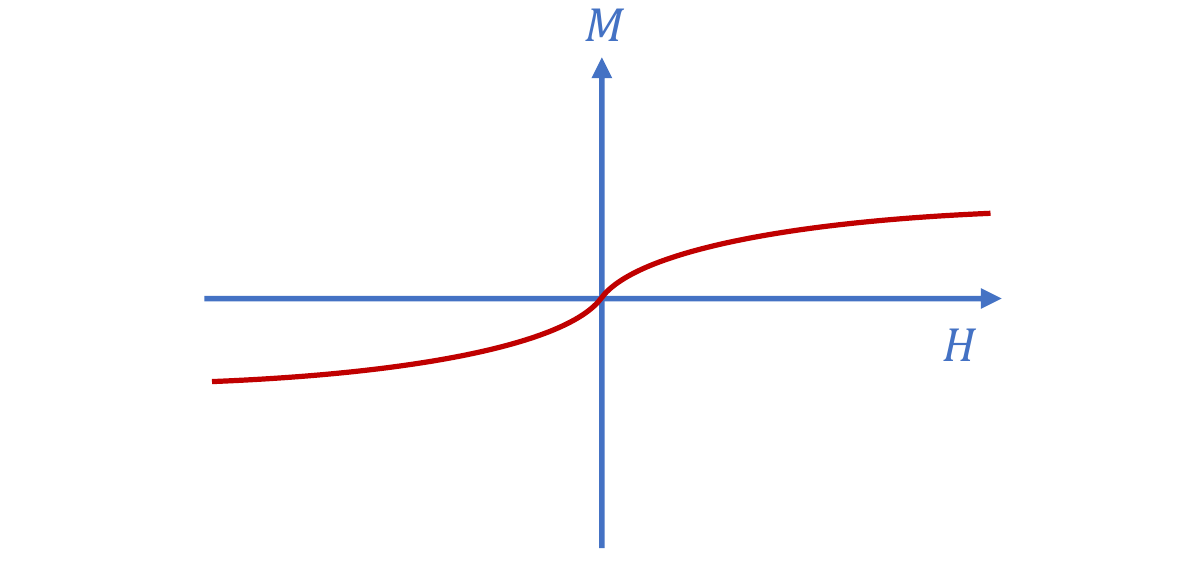}
\caption{\small Graphs of $M(H,T)$. Left: $T=T_{\tn C}$. Right: $T>T_{\tn C}$.}
\label{magnetiz-bc}
\end{center}
\end{figure}

There is a critical temperature, the Curie temperature of the material, $T_{\tn C}$, at which the property of spontaneous magnetization is lost. As one can see from Figure \ref{magnetiz-bc}, for $T=T_{\tn C}$ and $T>T_{\tn C}$, the magnetization goes to zero when the external field is removed, and so the material is {\it not} spontaneously magnetized above the Curie temperature.

The behaviour near the critical temperature is conveniently described by a {\it critical exponent}, $\b$:
\bea
M_0(T)\simeq(-t)^\b~,~~~~\mbox{as}~t\rightarrow0^-~,~~~~\mbox{where}~t:=(T-T_{\tn C})/T_{\tn C}~.
\eea

So far, I have discussed the ferromagnet in a quasi-phenomenological way, without the need to deploy a specific microscopic model. To calculate the critical exponents, the Ising model is often adopted, in which the molecules of the ferromagnet are thought of as lying on the sites of a regular lattice. Each molecule is assigned a magnetic moment (or `spin') that can take only two values, $+1$ and $-1$. The Ising model then assigns a specific Hamiltonian with two terms, one which models the interactions between neighbouring spins, and one which models the interaction between the spins and the external magnetic field. In what follows, we will not need the details: see e.g.~Binney et al.~(1992:~pp.~55-56, 61-65), Chandler (1987:~pp.~119-123), Baxter (1982:~pp.~14-15). 

Among the various techniques, used to calculate the critical exponents within such models, are: mean field theory, high-temperature expansions, and renormalization group theory. For the discussion here, I will follow the exact treatment of the two-dimensional Ising model (i.e.~the lattice is taken to be a two-dimensional square lattice), originally due to Onsager (1944).\footnote{I here follow the exposition in Baxter (1982: Chapter 7); see also Binney et al.~(1992:~Chapter 3).} The model is admittedly low-dimensional, and so it is only indicative of what happens in real materials: but it has the great merit of being exact, so that there is no question that, within the given idealizations, the emergence is ontological (cf.~Section \ref{mando}) rather than being a matter of mere coarse-graining.\footnote{The transfer matrix techniques, developed to solve this model, have been generalised to give treatments of more complex models (cf.~Baxter (1982:~Chapters 5-14)).}

In the two-dimensional Ising model, the critical exponent is $\b=1/8$. The spontaneous magnetization is calculated to be:
\bea
M_0(T)=(1-k^2)^{1/8}~,~~~~~~k:=1+t=T/T_{\tn C}~,~~~~~~0<k<1~.
\eea
Thus the spontaneous magnetization is highest, $M_0(0)=1$, at $T=0$, so that all the spins are aligned; and the magnetization monotonically decreases to zero, at the Curie temperature. Above the Curie temperature, it remains zero.

A quantity that is often introduced and that precisely characterises the dynamics of the Ising model is the {\it correlation length}, $\xi$. The correlation length is defined as the length scale that marks the exponential fall-off of the correlation function between two arbitrary spins on the lattice. Thus when the correlation length is of the same order as the lattice spacing, the values of two spins are correlated (as being, on average, both ``mostly plus'' or both ``mostly minus'') only if they are near-neighbours. The limit of very short correlation length $\xi$ corresponds to a disordered state, where the spins are randomly arranged with no correlations whatever. We expect this to happen at very high temperatures, since high temperature tends to randomize the spins. On the other hand, if the correlation length is very large (much larger than the lattice spacing), the system is highly ordered. We expect this to happen at the critical temperature, and indeed this property of an infinite correlation length is often `regarded as the hallmark of a critical point' (Baxter (1982:~p.~19)). In the two-dimensional Ising model discussed above, the correlation length near the Curie temperature is given by:
\bea
\xi=-{1\over\ln T/T_{\tn C}}\simeq{1\over1-T/T_{\tn C}}\rightarrow+\infty~~~~~~\mbox{as}~T\rightarrow T_{\tn C}^-~.
\eea
Thus the system is macroscopically {\it ordered}.\\

In characterising emergence, some authors have emphasised the mathematical properties of the model, especially the discontinuity in the magnetization and the infinite correlation length (see e.g.~Batterman (2002:~Chapter 4), who discusses `singular limits' and `singular asymptotics'). But these mathematical discontinuities and infinities---important as they may be---are not essential for emergence. After all, some of the emergent behaviour might be visible before we reach $T=T_{\tn C}$: for example, if finite-size effects or impurities smoothen the transition (cf.~footnote \ref{actex}).\footnote{For a discussion of ``emergence before the limit'', see Butterfield (2011a:~pp.~1069, 1089, 1098-1100)).}

My characterisation of emergence, rather than relying on specific mathematical properties of the models, focusses on the model's {\it change of interpretation} at or near the limit. 

Namely, at the critical temperature we have the appearance of order (i.e.~a macroscopic correlation length, much larger than the lattice distance $a$, $\xi\gg a$), and of spontaneous magnetization, i.e.~of a non-zero magnetization field $M_0(T)$ of the material when we remove the external magnetic field.

These properties of the ferromagnet at $T=T_{\tn C}$ are novel, relative to those at $T>T_{\tn C}$. Thus it is natural to compare these two models: namely, to take the Ising model at $T=T_{\tn C}$ to be the top theory, $T_{\sm t}$, and the Ising model at high temperature, $T>T_{\tn C}$, to be the bottom theory, $T_{\sm b}$.\footnote{We are interested in synchronic emergence, and so we take the two phases to be synchronic levels, related by the temperature gradient. But one might be interested in diachronic emergence: in which case one would consider gradually lowering the temperature in time (while at the same time turning the magnetic field higher or lower).} We get from $T_{\sm b}$ to $T_{\sm t}$ by lowering the temperature towards the Curie point, i.e.~taking $T\rightarrow T_{\sm C}^+$.\footnote{More explicitly, the emergence base is the collection of Ising models at temperature above the Curie temperature, $\{T_{\tn b}(T)\}_{T>T_{\tn C}}$. We then get the top theory, $T_{\sm t}$, by taking the limit, as in Eq.~\eq{parx0}. One also easily sees that the map link, thus defined, is non-injective, as it should be: for it maps models (i.e.~states and quantities) at any $T>T_{\tn C}$ to a single model (to its states and quantities) at the Curie temperature, i.e.~``it forgets about the temperature''. This is in agreement with the remarks in footnote \ref{basepoint}.\label{curiet}} 

These properties are elements of the domain of application of the theory at $T< T_{\tn C}$, but not at $T>T_{\tn C}$. Namely, it makes a difference whether we first interpret the bare model and then lower its temperature, or first lower the temperature and then interpret the bare model.\footnote{I am here adopting the talk of `bare theory', from Section \ref{ern}, for formal, i.e.~uninterpreted, models.} In the latter case, the interpretation will contain in its range elements, such as (i) a macroscopic magnetization of the material, and (ii) a macroscopic correlation length, that are not part of the interpretation of the model above the critical temperature. 

Thus, taking the bottom theory, $T_{\sm b}$, to be the Ising model above the critical temperature, and the top theory, $T_{\sm t}$, to be the Ising model at critical temperature, we see that the two models are mathematically linked,\footnote{Due to discontinuities and divergences involved in the use of limits, this link is not a case of reduction. See the discussion at the end of Section \ref{nornc}.} and yet their interpretations are different, as in Figure \ref{4leginterp}. Namely, the interpretation of the top theory ascribes properties to the system (viz.~macroscopic spontaneous magnetization and correlation length) that do not apply to the bottom system. 

The example in the next Section will show in detail how to exhibit the formal requirement of non-commutativity (which the previous paragraph has described, in words).

\section{Second Case Study: Masslessness in Relativistic Mechanics}\label{mmpp}

In recent theoretical physics there is a pervasive idea that, for theories with external parameters, and taking some of those parameters to special values, one gets to a ``critical, or massless, regime'' of the theory. The idea is that, as the physical length scale of the massive theory disappears, the theory is in a scale-invariant, or massless, regime. This is the idea of conformal fixed points in both statistical mechanics and in quantum field theories. Thus emergence is often associated with the appearance of such a massless or scale-invariant regime within a massive theory: for it is commonly agreed that there is an important sense in which the massless properties that arise in the special regime---such as, for example, symmetries---are emergent. 

In this Section, I will illustrate some of these ideas in a simple model: namely, the emergence of a massless free classical point-particle, within the theory of relativistic classical mechanics for massive particles. It will concern the emergence of properties within a unified theoretical framework.\footnote{For an example of emergence between different theories, see Section \ref{casemm}.} Since the example is simple, a short account can be more detailed than that of the previous Section: thus exhibiting all the necessary ingredients of my account of ontological emergence. I describe, in Section \ref{bpp}, the bottom theory, of which I will take the $m\rightarrow0$ limit. In Section \ref{encai}, I show that this illustrates emergence. In Section \ref{massmass}, I discuss in more detail the differences between massive and massless particles. 

For simplicity of the presentation, I will take the bare theory to be just the equations of motion, derived from a Lagrangian. This is justified for our purpose in this Section because, in a classical theory of the kinds I will discuss, the equations of motion allow us to construct all the states and quantities of the theory.

\subsection{The bottom theory $T_{\sm b}$ for the point-particle}\label{bpp}

The equations of motion for a free classical point-particle of mass $m$ in Minkowski space are our bottom theory, $T_{\sm b}$. They are as follows:\footnote{These equations can be derived from a Lagrangian, which is discussed in detail in De Haro (2019a:~\S3.1).}
\bea
{\dd\over\dd\s}\left({1\over e}\,{\dd x^\m\over\dd\s}\right)&=&0\label{firsteq}\\
\eta_{\m\n}\,{\dd x^\m\over\dd\s}\,{\dd x^\n\over\dd\s}+e^2\,m^2&=&0~.\label{lightcone}
\eea
The variable $e(\s)$ is a real, positive-definite, Lagrange muliplier, introduced in order to deal with the massive and massless cases using a unified theoretical framework. Here, the Minkowski metric is $(\eta_{\m\n})=\mbox{diag}(-1,1,1,1)$, and $\s$ is an affine parameter, viz.~a monotonically increasing parameter, along the particle's world line. The equations of motion are invariant under reparametrisations of the affine parameter $\s$, which implies that $e(\s)$ can be transformed away (i.e.~set equal to one, if $m\not=0$) by a choice of the affine parameter, $\s$.  

The equations of motion, Eqs.~\eq{firsteq}-\eq{lightcone}, actually form a one-parameter family of theories, where the parameter is the mass, $m$ (though I will keep simply referring to Eqs.~\eq{firsteq}-\eq{lightcone} as `the theory'). Thus, we have a theory for each value of $m$, which I will denote as $T_{\sm b}(m)$, to indicate that the theory ``fixes the mass''. Accordingly, I take the mass to also be in the domain of the interpretation map, which is evaluated on $T_{\sm b}(m)$.

In the massive case ($m\not=0$), it follows from Eq.~\eq{lightcone} that the proper time is $\dd\t=mc\,e(\s)\,\dd\s$, so that Eq.~\eq{lightcone} simplifies to the condition that the velocity 4-vector is time-like. 

When $m=0$, Eq.~\eq{lightcone} is the condition that the 4-velocity vector lies on the light-cone.\footnote{One may wonder whether it makes sense to take the $m\rightarrow0$ limit for a {\it single,} free particle. Regardless of one's answer to this: the simple point-particle model considered here of course does not stand on its own, but is to be coupled to further physics, which will determine the scale relative to which the mass is taken to zero. In Section \ref{massmass}, we will consider collisions of two particles.}

\subsection{Emergence as non-commutativity of linkage and interpretation}\label{encai}

In this Section, I discuss the emergence diagram, in Figure \ref{4leginterp}, for the theory of the massive particle, Eqs.~\eq{firsteq}-\eq{lightcone}. I will do this by constructing the analogue of Eq.~\eq{parx}, i.e.~the lack of commutativity, for a theory with a continuous parameter.

The point about emergence is this: as long as $m$ is non-zero, the interpretation of the equations of motion, Eqs.~\eq{firsteq}-\eq{lightcone}, is the familiar one. Namely, $x^\m(\t)$ gets interpreted as `the position of a massive particle, with mass $m$ [some number fixed by the theory], moving freely in Minkowski space, as a function of the particle's proper time'. The interpretation is {\it the same} for all nonzero values of $m$, i.e.~it is the same map: for, as we saw in Section \ref{nornc}, the interpretation map depends on $m$ through the theory's dependence on it (here, $m$ plays the role of the parameter $x$, in Eqs.~\eq{parx0}-\eq{parx}).

In the $m=0$ case, the interpretation of $x^\m(\s)$ is: the position of a massless particle moving at the speed of light in Minkowski space, as a function of the affine parameter of its worldline. 

The main difference between the two interpretations is in how `massive' and `massless particle' are construed. I will discuss this in more detail in Section \ref{massmass}.\\ 

\begin{figure}
\begin{center}
\includegraphics[height=4cm]{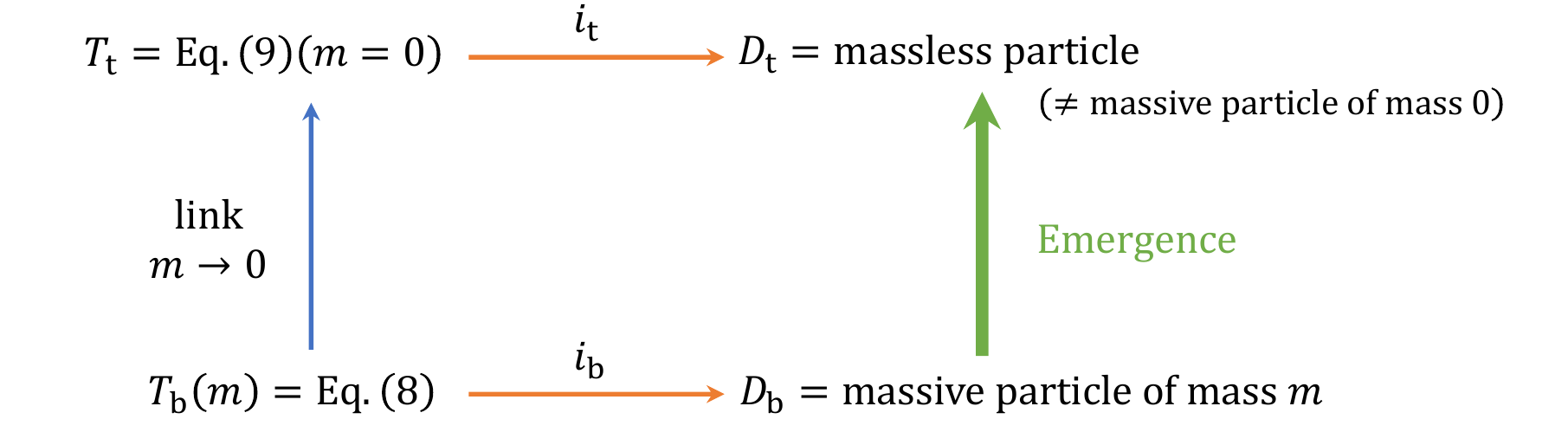}
\caption{\small Emergence of the massless particle, analysed as the lack of commutativity between the linkage map, $m\rightarrow0$, and the interpretation: $i_{\sm t}\circ\,\mbox{link}\not=i_{\sm b}$.}
\label{massless}
\end{center}
\end{figure}

Here is what it means for the diagram in Figure \ref{4leginterp}, according to the interpretation in Eq.~\eq{parx} for a one-parameter family of theories, to describe emergence: the linkage map is the process of taking the limit $m\rightarrow0$. So, we start with the bottom theory, for any non-zero $m$:
\bea\label{tb}
T_{\sm b}(m)=\mbox{`the theory (Eqs.~\eq{firsteq}-\eq{lightcone}) for the particle of mass $m$'}~,
\eea
where such qualifications as `free point-particle' are, for simplicity of presentation, already included in what `the theory' is, namely in Eqs.~\eq{firsteq}-\eq{lightcone}. Taking the limit $m\rightarrow0$ as the linkage map, we end up with the top theory:
\bea\label{tt}
T_{\sm t}:=\mbox{link}\,(T_{\sm b})=\lim_{m\rightarrow0}T_{\sm b}(m)=\mbox{`the theory (Eqs.~\eq{firsteq}-\eq{lightcone}) for the particle of mass 0'}.
\eea
The limit is smooth, and it is represented by the leftmost arrow in Figure \ref{massless}. 

Emergence occurs because {\it what is described}, in the limit, is different---indeed novel, as represented by the rightmost arrow in Figure \ref{massless}. We have all the tools in place; so that this is now straightforward to show. 

First, the interpretation of the bottom theory, Eq.~\eq{tb}, has as its range (in a simplified form that suffices for our discussion):
\bea\label{itb}
D_{\sm b}:=i_{\sm b}(T_{\sm b}(m))=\{\mbox{a free, massive point-particle of mass $m$}\}~.
\eea
This reflects the fact that, as I mentioned in Section \ref{bpp}, the interpretation is a function of the mass, and the domain of application may contain particles of any mass. 

The interpretation of the top theory, Eq.~\eq{tt}, has as its range:
\bea\label{itt}
D_{\sm t}:=i_{\sm t}(T_{\sm t})=\{\mbox{a free, massless point-particle}\}\overset{{\sm{Eq.}}~\eq{tt}}{=}i_{\sm t}\,(\mbox{link}\,(T_{\sm b}))~.
\eea
But if we go along the lower horizontal arrow, $i_{\sm b}$, in Figure \ref{massless}, and then take the value $m=0$ in Eq.~\eq{itb}, we do {\it not} end up at $D_{\sm t}$ (i.e.~Eq.~\eq{itt}). For the range of the interpretation $i_{\sm b}$ is incorrect for a massless particle, namely:
\bea\label{ibm0}
D_{\sm b}\,|_{m=0}=i_{\sm b}(T_{\sm b}(m))|_{m=0} \overset{{\sm{Eq.}}~\eq{itb}}{=}\{\mbox{a free, massive point-particle of mass 0}\}~.
\eea
This is different from the correct (because more accurate) interpretation: namely, Eq.~\eq{itt}. 

Thus the two interpretation maps, $i_{\sm t}$ and $i_{\sm b}$, map to different domains, even though both have $m=0$. The underlying point is that, as I will argue in Section \ref{massmass}:
\bea\label{massmassive}
\{\mbox{a free, massless point-particle}\}\not=\{\mbox{a free, massive point-particle of mass 0}\}~,
\eea
i.e.~the massless particle is different from the massive particle `with the mass set to zero' (and the former is also not contained in the latter as a proper subset). Thus, we have proven the conditions, Eqs.~\eq{dnotd} and \eq{itnotib}, that the linkage and interpretation maps do not commute. The interpretation of $T_{\sm t}$ as describing a {\it massless} particle is a novel interpretation, relative to the massive particle interpretation. This is why there is ontological emergence.

Notice that there is no requirement that the top and bottom domains, $D_{\sm t}$ and $D_{\sm b}$, must ``greatly differ'', or that their differences must be striking, in order for there to be emergence; see my comments about this at the end of Section \ref{comparrw}. There is simply emergence if they differ. But agreed: the more they differ, the greater the novelty. 

\subsection{Massive vs.~massless particles}\label{massmass}

In this Section, I argue that classical massive and massless particles are qualitatively different (thus substantiating Eq.~\eq{massmassive}). Indeed, although they are both described by the equations of motion, Eqs.~\eq{firsteq}-\eq{lightcone}, their properties are distinct. I will discuss three related properties that differ between the massless and massive cases and that, together, characterise a massless particle:\\

(i)~~{\it Timelike vs.~null geodesics.} The first difference is in the geodesics, namely in the time-like condition vs.~the null condition, Eq.~\eq{lightcone}. In terms of the four-momentum vector, $p:=m\,\dot x$, these conditions are: $p^2=-m^2c^2$ in the massive case, vs.~$p^2=0$ in the massless case. 

In the limit $m\rightarrow0$, the massive equation, $p^2=-m^2c^2$, of course reproduces the massless equation, $p^2=0$. And thus the timelike geodesics converge to null geodesics. Yet this reduction does not prevent emergence, for it is reduction in the bare theory and not in the domain of application. Namely, the property of `being a timelike geodesic' still figures in the interpretation of the bottom massive theory, even though we set $m=0$. Thus the interpretation does not fit with the idea of a `massless particle', in other words: the range of the interpretation is empty if the interpretation is an extension, i.e.~$i_{\sm b}(T_{\sm b}(m))|_{m=0}=\O$.\footnote{This concurs with Norton's (2012: pp.~208-210; 2016:~\S3.1) idea that limits of theories sometimes do not describe physical systems, real or fictitious, because they ascribe to them {\it contradictory properties.} More details follow below, in an example of a collision.} One might at first sight think that this is only a matter of words, since it is clear from a Minkowski diagram that the $m\rightarrow0$ limit of a timelike geodesic is a null geodesic. But it is, in fact, not a matter of words: for there are also consequences for the interpretation of the massive theory, of sending $m\rightarrow0$, that plainly {\it contradict} the interpretation of the massless theory.

One such consequence is the absence of a rest frame for a massless particle, while there always exists one for a massive particle. 

\begin{figure}
\begin{center}
\includegraphics[height=5cm]{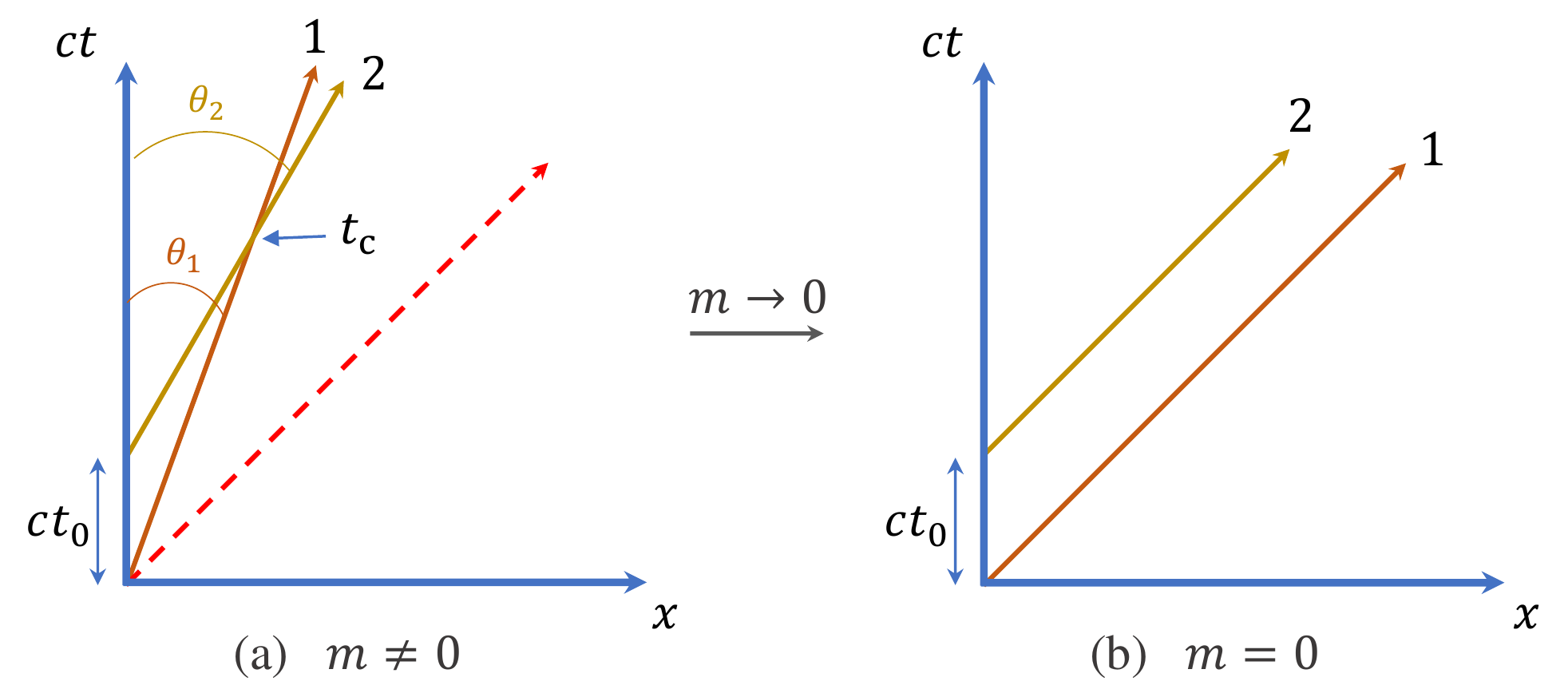}
\caption{\small (a) Two particles with constant speeds, $v_1/c=\tan\th_1$ and $v_2/c=\tan\th_2$, colliding at $t=t_{\sm c}$. (b) Massless limit, $v_1=v_2=c$: the particles {\it do not} collide at any finite time. As $m\rightarrow0$, the lines in (a) are continuously deformed to the lines in (b) (but the {\it crossing point} is pushed to infinity!).}
\label{collision}
\end{center}
\end{figure}

Another interesting consequence regards the {\it  occurrence of events,} e.g.~collisions. Consider two particles with mass $m$, travelling at different speeds in the same forward direction, but separated by an initial interval of time $t_0$, i.e.~particle 2 is sent from the same location, but at a time $t_0$ later than, particle 1: see Figure \ref{collision}(a). Take the second particle to travel faster than the first, so that they are due to collide when the second overtakes the first. But if we take the massless limit, these two particles will {\it never} collide, for they will both travel at the speed of light, and they will keep their mutual initial separation (see Figure \ref{collision}(b)). In other words,  an event that is possible, according to $i_{\sm b}$, at non-zero $m$ under the given initial conditions (namely, `the particles will collide within finite time') becomes impossible, according to $i_{\sm t}$, in the limit $m\rightarrow0$, under the corresponding initial conditions (namely, `the particles will not collide within any finite time'). 

To illustrate this distinction by a model, consider again Figure \ref{collision}(a). The particles move along straight lines, at constant speeds $v_1$ and $v_2$, respectively, where $0<v_1<v_2<c$. Their trajectories are given, respectively, by:
\bea\label{trajm}
x_1(t)&=&v_1\,t\nn
x_2(t)&=&v_2\,(t-t_0)~.
\eea
The time, $t_{\sm c}$, at which the two particles collide is calculated by setting: $x_1(t_{\sm c})= x_2(t_{\sm c})$. The result is: 
\bea\label{collt}
0~<~~t_{\sm c}={c\over v_2-v_1}~t_0~~<~\infty~.
\eea

By taking the $m\rightarrow0$ limit of the trajectories, one finds that the two particles travel at the speed of light, i.e.~$v_i\rightarrow c$ ($i=1,2$), as in Figure \ref{collision}(b): namely, the two particles' trajectories are taken to lie on separate, right-moving, null lines. The time of collision, Eq.~\eq{collt}, gets ``pushed to infinity'', i.e.~$t_{\sm c}\rightarrow\infty$: for, in the limit, the speed of the first particle approaches the speed of the second particle from below, and they both travel at the speed of light (so that the condition $v_1<v_2$ is {\it violated!}).\footnote{Malament (1986:~pp.~192-193) discusses a somewhat similar case, namely the geometric (Cartan) version of Newtonian gravitation, as the $c\rightarrow\infty$ limit of general relativity. In this limit, the light-cones ``open up'' and become flat, thus violating the four-dimensional general covariance of the theory.}\\

There is an insightful way to understand why the maps fail to commute. Namely, we use the notion of a model of a theory, as being part of the interpretation map: we construct our interpretation map of $T_{\sm b}$ by constructing models like Eqs.~\eq{trajm}-\eq{collt} and Figure \ref{collision}(a). The point, then, is that these massive models instantiate concepts such as `timelike geodesic', `particle's rest frame', and `finite collision time': but the limit $m\rightarrow0$, Figure \ref{collision}(b), does not instantiate any of these notions. Thus there are no models of $T_{\sm b}$ of this kind in the limit. 

For example, in the particle collision model it is true that, for any $m\not=0$, the collision time is finite, as stipulated by Eq.~\eq{collt}. But the limit $m\rightarrow0$ {\it violates} this condition, because it requires that the two particles travel at the same speed, viz.~$v_1\rightarrow v_2^-$, so that $t_{\sm c}\rightarrow\infty$: and so, there is in the limit {\it no model} of $T_{\sm b}$ with a collision of this kind. 

Notice that, while this is closely related to Norton's idea of `ascribing contradictory properties to the limit system', in this case there {\it is} a limit system, namely the one that can be constructed by interpreting the massless theory directly, through $i_{\sm t}$ (cf.~Eq.~\eq{itt}): so that we have a genuine case of idealisation, along the lines discussed in Section \ref{nnr}. This gives rise to a limit system that is arbitrarily close to the massive one (cf.~Eq.~\eq{itb}) in some aspects, but not in its novel aspects: and it is {\it not} obtained by taking the limit of the massive system.

Summarising, the interpretation of the bottom theory, $i_{\sm b}$, assigns contradictory properties to the particle in the $m\rightarrow0$ limit: for it still contains concepts that it inherits from the massive case and which are contradicted by the massless limit, and which no longer refer. This is why, to get the top theory's interpretation, $i_{\sm t}$, one needs to go back to the $m\rightarrow0$ limit of the bare theory, i.e.~go back to $T_{\sm t}$, and build (at least some aspects of) its interpretation again. The interpretation of the limit theory is not the limit of $i_{\sm b}$.\\

(ii)~{\it Symmetries of the solutions.} The states of the massive and the massless particles have different symmetries. To show this, we consider a generic solution of the equations of motion, Eqs.~\eq{firsteq}-\eq{lightcone}, and construct the subgroup of the Poincar\'e group leaving this solution invariant. Wigner (1939:~p.~50) dubbed this the `little group'. 

First consider a massive particle, with given four-momentum vector $p=m\,\dot x$, obtained by solving the equations of motion, Eqs.~\eq{firsteq}-\eq{lightcone}. To find the subgroup of the Lorentz transformations leaving this vector invariant, go to the particle's rest frame, where $(p^\m)=mc\,(1,{\bf 0})$. The subgroup of the Lorentz group leaving this vector invariant consists of the spatial rotations, plus in addition time reversal symmetry. This group is isomorphic to O(3). 

In the massless case, there is of course no rest frame for the particle: but we can orient our coordinate system in such a way that the symmetries are easy to see. The following is a useful choice: $(p^\m)=p\,(1,0,0,1)$, where $p$ is the particle's momentum. The group preserving this vector is the group of Euclidean motions of the plane, E(2) $\cong$ ISO(2) (see e.g.~Gilmore (2008:~\S13.3)). It contains three generators: one rotational (corresponding to rotations around the $x^3$-axis) and two translational, along the $x^1$ and $x^2$-directions.\footnote{For a discussion of their physical interpretation, see Han et al.~(1981:~\S II).} Therefore, the little group is isomorphic to E(2). 

Notice that E(2) is {\it not} a subgroup of O(3): it has different generators, and so the symmetries of the two cases are different, which expresses the {\it novelty} of the symmetry group in the massless case, compared to the massive case.\footnote{O(3) and E(2) are related by group contraction (see In\"on\"u and Wigner (1953), Gilmore (2008:~\S13.3.1)), which has a geometric interpretation in terms of ``projection to a flat surface, around the North pole of the sphere'' associated with O(3), i.e.~a ``large radius'' limit. See Kim~(2001:~\S4).}\\

Recall, from Section \ref{whichs}, the three possible cases, (1)-(3), for comparison between the idealising and the target systems. In so far as the massless theory is taken to give a description of photons or other massless particles, we have a case of (2), i.e.~the top theory gives a good description of massless particles, while the bottom theory does not. The usefulness of emergence is then that it exposes the interpretative differences between the massless and the massive cases, while it gives a unified description of the two cases through the linkage relation.\footnote{One might also want to apply the massive theory to describe massive, but very light, particles. Pitts (2011:~p.~277) argues that the massless case is empirically distinguishable from any {\it particular} massive theory, but only by observations probing length scales of the order of the inverse mass. Pitts goes on to argue that `there exists a range of sufficiently small photon masses such that the massive... theories are empirically indistinguishable from the massless theory'.} 

\section{Third Case Study: Riemann Surfaces in Random Matrix Models}\label{casemm}

In this Section, I will briefly introduce random matrix models and explain the sense in which, in these models, a space is emergent from a discrete structure.\footnote{Some philosophical questions about the use of these models in quantum gravity theories are discussed in De Haro and De Regt (2018).}

Let me say a few words about the philosophical interest of random matrix models. Their use is ubiquitous in many areas of physics and beyond: ranging from statistical mechanics to Wigner's ground-breaking work on the energy levels of heavy nuclei, to quantum chaos in billiards and disordered systems, integrable systems, crystal growth, number theory, the channel capacity of digital communication systems, and RNA folding in biology.\footnote{There is even an Oxford Handbook of Random Matrix Theory: Akemann et al.~(2011).}  Since random matrix models describe complex theories in simplified settings while retaining much of the characteristic physical behaviour and properties---such as universality, phase transitions, renormalization, and quantum-to-classical transitions---they can be of as much use to philosophers as they are to physicists: for they give ``precise conceptual laboratories'' in which to test ideas. Random matrix models also play a central role in several approaches to quantum gravity that display emergence of space and-or time: and it is from this perspective that I introduce them here.

Random matrix models are, roughly speaking, quantum theories whose states are specified by an $N\times N$ matrix $\F$. A physical way to think about these matrices---and this has been one of the main motivations for studying random matrix models in high-energy physics---is as simplified versions of quantum fields with colour charges for the strong nuclear force, as in quantum chromodynamics.\footnote{See 't Hooft (1974), Br\'ezin et al.~(1978).} The matrix structure of the field reflects its non-abelian colour charges. This is a simple model of colour charge, because the field is a constant matrix, without any space or time dependence. The bare theory (in the sense of Section \ref{ern}) has a Lagrangian, but without a kinetic term. The potential is (the trace of) a polynomial, $W$, in the field, $\F$. Finally, one constructs a suitable integration measure over the possible values of the matrix entries.\footnote{The {\it quantities} of this theory (cf.~Section \ref{ern}) are the set of all the polynomials of the matrix $\F$. The {\it rules for evaluating physical quantities} on the states are quantum-theoretic: a matrix integral with its measure (called the `partition function', on analogy with quantum field theory), weighted by the exponential of the potential, and with appropriate insertions of the physical quantities one wishes to evaluate. There is no {\it dynamics} in the usual sense, because there is no time.} 

One can intuitively think of these matrices as charged fields, in the limit in which the dynamics is dominated by the potential, and the kinetic term is negligible: i.e.~the field is interacting, but slowly varying in space and time. \\

I will next describe the linkage map and then discuss emergence. The linkage map is the 't Hooft approximation: namely, taking $N$ to be large, the coupling constant $g$ to be small, while keeping the combination $\m:=g^2N$ fixed (which becomes the effective coupling constant). Thus we consider a sequence of random matrix models, with matrices of increasing rank $N$.\footnote{Renormalization of a subset of the quantities is also required. See De Haro (2019b:~Appendix B).}

The quantities (such as the free energy) of the random matrix model can be calculated using the saddle point approximation to the path integral. These are best written in terms of the $N$ eigenvalues, $\l_i$, of the matrix $\F$; and this calculation of quantities leads to a ``zero-force'' condition, which we can think of as the random matrix model's equation of motion, in the semi-classical limit introduced by the saddle point approximation:
\bea\label{leom}
W'(\l_i)-{2\m\over N}\,\sum_{j\not=i}{1\over \l_i-\l_j}=0~~~~.
\eea
The eigenvalues of the matrix, $\l_i$, can here be thought of as the colour charges of the field, namely as states in colour space. The first term in Eq.~\eq{leom} is the gradient of the classical potential $W$, i.e.~the ``force'' (which is a polynomial of degree $n$), and the second term is the quantum contribution of the measure factor in the path integral, which takes the form of a ``Coulomb repulsion'' between distinct pairs, $(i,j)$, of colour charges. \\

The behaviour of the solutions to the equation of motion, Eq.~\eq{leom}, is as follows. We consider the 't Hooft approximation, where (as mentioned) we will increase $N$ but keep the value of $\m:=g^2N$ fixed (and so, we will take the coupling constant $g$ to decrease as $N$ increases). This means that we consider a sequence of models with increasing number of colours (in quantum chromodynamics, $N$ is fixed to $N=3$), where $\m$ is the effective coupling constant of the interactions. This approximation of large $N$ is reminiscent of a ``thermodynamic limit'', and it turns out to give interesting physics.

First, for any {\it finite} value of the number of colour charges, $N$, and for small $\m$, the equation of motion, Eq.~\eq{leom}, implies that $W'(\l_i)\simeq 0$, and so the $N$ eigenvalues, $\l_i$, are distributed among the $n$ extrema of the classical potential. Since $n<N$, there are double points where the eigenvalues are degenerate.

Second, for large $N$ and large $\m$, the Coulomb repulsion (the second term in Eq.~\eq{leom}) becomes relevant, and most eigenvalues repel each other; so that the degeneracies disappear, and the line segments between the $n$ extrema get filled by eigenvalues, because there are many more eigenvalues than extrema. It turns out that in the limit $N\rightarrow\infty$, keeping $\m$ fixed, the line segments are filled by eigenvalues. 

Upon solving Eq.~\eq{leom} for $N\rightarrow\infty$, the solution can be summarised by a single complex function, which encodes the ``location of the eigenvalues'' in the complex plane parametrised by $x\in\mathbb{C}$ (in the $N\rightarrow\infty$ limit, the eigenvalues lie on a continuum, as just argued!). It is given by:
\bea\label{yofx}
y=\sqrt{\prod_{\b=1}^{2n}(x-a_\b)}~.
\eea
$y$ is called {\it the spectral curve of the matrix model}.
One recovers from $y$ the density of the eigenvalues on the complex plane, which in turn allows one to calculate the physical quantities of the random matrix model, such as the free energy.\\ 
\\
The central result needed for emergence can be summarised in the following:\\
\\
{\bf Proposition}.\footnote{The proposition is proven in De Haro (2019b:~\S2.3.2). See also Dijkgraaf and Vafa (2002:~\S3.2) and Eynard et al.~(2015:~\S1.3.2).} {\it The spectral curve of the random matrix model, Eq.~\eq{yofx}, is a compact, hyperelliptic Riemann surface, $\Sigma_g$, of genus $g=n-1$.}\footnote{One can see from Eq.~\eq{yofx} that $y$ has branch cuts between the even and odd $a_\b$'s (where $\b=1,\ldots,2n$). These branch cuts can be made single-valued by taking the double cover of the complex plane, with appropriate identifications---thus obtaining the $n-1$ handles of the Riemann surface, from the $n$ cuts.}\\

The proposition allows us to describe the 't Hooft approximated theory, $T_{\sm t}:=\mbox{link}\,(T_{\sm b})$, as a triple. $T_{\sm t}$ is a classical theory, whose {\it states} are configurations of a Riemann surface of genus $g$, $\Sigma_g$, with a local chart $(x,y)$ satisfying Eq.~\eq{yofx}, for given coefficients $\{a_\b\}_{\b=1}^{2n}$ (whose values are determined by the classical potential of $T_{\sm b}$). The {\it quantities} (for example, the free energy) are now obtained from the one- and two-forms with which this Riemann surface is equipped. The dynamics is Eq.~\eq{yofx} itself, for specific choices of coefficients $\{a_\b\}_{\b=1}^{2n}$ (and, as in the bottom theory, there is no time: though there is now {\it space}, viz.~$\Sigma_g$ itself!). 

To understand the physics of this theory, $T_{\sm t}$, an analogy with general relativity may be useful. There, the models, $({\cal M},g,\{\F\})$, are triples of: (i) Lorentzian spacetimes, (ii) endowed with a smooth metric, and (iii) matter fields. In the case of $T_{\sm t}$, we have the following: 

(i)~There is a compact two-dimensional Euclidean space, $\Sigma_g$, with genus $g$ (recall that the topology of a compact two-dimensional surface is completely determined by its genus). 

(ii)~The space is not endowed with a metric, but it is endowed with a complex structure, Eq.~\eq{yofx}, and a set of charts, $(x,y)$. 

(iii)~We have one- and two-forms, on this Riemann surface, that give the quantities of the theory (and are analogous to the ``matter fields'', though they are really geometrical quantities). \\

Thus the state spaces describe very different kinds of objects: colour charges (with no space at all!) vs.~a curved two-dimensional surface, viz.~a Riemann surface $\Sigma_g$. More explicitly: taking the limit $N\rightarrow\infty$ of the bottom theory's interpretation (i.e.~the right-hand side of Eq.~\eq{parx}, with $x=1/N$), we get the interpretation of a state as `an infinite set of colour charges that are weakly interacting' (since we also keep $\m=g^2N$ fixed, and thus send $g\rightarrow0$). This differs from the top theory's correct interpretation (the left-hand side of Eq.~\eq{parx}), as `the configuration of a Riemann surface'. Thus the domains of application are different, $D_{\sm b}\not=D_{\sm t}$. Then also: $i_{\sm t}\,\circ\,\mbox{link}\not=i_{\sm b}$ (and it is also obvious that $D_{\sm t}~\,/\!\!\!\!\!\!\subset D_{\sm b}$).

This establishes my earlier statement, that first taking the 't Hooft approximation and then interpreting, or taking the sequence of the interpreted theories (i.e.~first interpreting and then taking the linkage relation), lead to different results. Namely, a Riemann surface is emergent from a set of colour charges.

\section{Conclusion}\label{conclusion}

In this paper I have analysed ontological emergence as novelty of reference, in terms of two conjuncts: a formal, i.e.~not interpretative, condition---linkage---and an interpretative condition---ontological novelty, which is a difference in intension, and sometimes also extension. One typical case of emergence is when the theories have the same extension but different intensions. I illustrated the idea of ontological emergence as the non-commuting diagram in Figure \ref{4leginterp}, namely the failure of the interpretation to commute with the linkage map, Eq.~\eq{itnotib}.

I illustrated ontological emergence in the examples of the emergence of spontaneous magnetization in a ferromagnet, the emergence of masslessness (or of a massless regime) for classical point-particles, and the emergence of space in a random matrix model. We saw that the interpretation of the massless particle theory is different from the interpretation of the massive particle theory with the mass set to zero. Namely, there are differences between their causal properties and symmetries, as well as in their predictions about whether certain collisions can take place. In the case of random matrix models, the bottom theory describes a set of colour charges, while the top theory describes the geometry of a two-dimensional surface. On the other hand, I discussed the example of phonons as a case of epistemic emergence.

My account of novelty is logically weak in that {\it any difference} in the description (epistemic novelty) or in the domain of application (ontological novelty) counts as novelty. This is as it should be: for `novelty' does not, by itself, carry a connotation of scientific or philosophical importance. And it is also {\it metaphysically} weak. For I have focussed on the basic question of when we are entitled to claim that ontological emergence obtains---a claim on which I have argued, in Section \ref{fdev}, that any interpretative strategy must agree, except for scepticism and radical forms of instrumentalism: which will dismiss ontology from the outset (although it is hard to do so consistently!). Further work, on the metaphysics of emergence, will have to further characterise the kinds of ontological novelty which arise in each case---namely properties, individuals, causal powers, etc. 

The account suggests various types of ontological emergence worth studying in more detail. One first distinction is between cases of emergence where the intensions of the two theories differ but their extensions are the same, vs.~cases where both the intensions and the extensions are different (see Section \ref{oee}). Another important distinction is whether the more accurate description of the target system is that of the bottom theory vs.~that of the top theory (see Section \ref{whichs}).

Unlike other accounts, which seek the mark of emergence in some technical property of bare theories (see (I) below), my explication formalises a conception of emergence that is intrinsically {\it interpretative} (which should not be confused with: subjective, or arbitrary!). The mark of novelty is the non-meshing of the linkage map with the interpretation. The main idea illustrated in Sections \ref{imct}, \ref{mmpp}, and \ref{casemm} was that the interpretation of the bottom theory differs from the interpretation of the top theory. Typically, the linkage map gave rise to an interpretation with properties that do not fit the top theory: as having zero magnetization and a disordered state with finite correlation length, as opposed to a `macroscopic magnetization' or `infinite correlation length'. Or as a `massive particle with mass 0', which differs from a genuine `massless particle' interpretation. Or as `an infinite set of colour charges' vs.~`a Riemann surface'.

I close with three further comments characterising the properties of the emergence relation which we have obtained:

(I)~~{\it ``Non-singular''.} We saw that the limits in Sections \ref{mmpp} and \ref{casemm} are not singular, in the mathematical sense: {\it pace} various philosophers' emphases.\footnote{Singular limits are essential for e.g.~Rueger's (2000:~p.~308) and Batterman's (2002:~pp.~80-81, pp.~1424-126) senses of emergence. On the other hand, Butterfield (2011a:~pp.~1073-1075) has argued that singular limits are not essential to have emergence.} In contrast with the well-rehearsed case of Section \ref{imct}, my examples from Sections \ref{mmpp} and \ref{casemm} show that there are no singularities, in the mathematical sense, in those cases: for all limits are smooth. 

(II)~~{\it ``Top-accuracy''.} A related aspect of emergence that the examples in Sections \ref{imct} and \ref{mmpp} illustrate is that the bottom theory does not always give a better (in the sense of: more accurate, both quantitatively as well as conceptually) description of the target system than the top theory that is being compared to. This is because, for any non-zero value of the mass, the bottom theory's interpretation does not give the correct properties of a massless particle, in particular: its geometric properties and its symmetries. (And likewise for the Ising model at different temperatures).

(III)~~{\it Very different interpretations and theories.} While the interpretations of the bottom and top theories and models involved in Sections \ref{imct} and \ref{mmpp} are ``of the same kind'', i.e.~both are about the Ising model or about point-particles, the two interpretations in Section \ref{casemm} are very different: and so, we get emergence between very different theories. Thus this is a good example of how a rich geometrical structure (a Riemann surface, equipped with its complex structure and set of differential forms on it) can be emergent with respect to a discrete set of colour charges, with no geometry at all (more details on this in De Haro (2019b:~\S4)).

\section*{Acknowledgements}

I thank Elena Castellani, Peter Kirschenmann, F.~A.~Muller, Hans Radder, and especially Jeremy Butterfield for comments on the paper and insightful discussions. I also thank two anonymous referees of their comments. I thank audiences at the Institut des Hautes \'Etudes Scientifiques (IHES), at the Inter-University Centre Dubrovnik, at Foundations 2018 in Utrecht, and at the Boston Colloquium for Philosophy of Science. This work was supported by the Tarner scholarship in Philosophy of Science and History of Ideas, held at Trinity College, Cambridge.

\section*{References}
\addcontentsline{toc}{section}{References}

\small

Akemann, G., Baik, J., and Di Francesco, P.~(2011). ``The Oxford Handbook of Random Matrix Theory''. Oxford: OUP. \\
\\
Anderson, P.~W.~(1972). `More is Different'. {\it Science}, 177 (4047), pp.~393-396.\\
\\
Anderson, P.~W.~(1989). `Theoretical Paradigms for the Sciences of Complexity'. In {\it A Career in Theoretical Physics}, Anderson, P.~W., 2005, Wold Scientific, 2nd edition.\\
\\
Batterman, R.~(2002). {\it The Devil in the Details}. Oxford University Press.\\
\\
Baxter, R.~J.~(1982). {\it Solved Models in Statistical Mechanics}.  Academic Press.\\
\\
Bedau, M.~A.~(1997). `Weak Emergence'. {\it Philosophical Perspectives}, 11, pp.~375-399.\\
\\
Bedau, M.~A.~and Humphreys, P.~(2008). `Emergence: Contemporary Readings in Philosophy and Science.' Cambridge, MA: The MIT Press.\\
\\
Binney, J.~J., Dorwick, N.~J., Fisher, A.~J., Newman, M.~E.~J. (1992). {\it The Theory of Critical Phenomena. An Introduction to the Renormalization Group}.  Oxford: Clarendon Press.\\
\\
Black, R.~(2000). `Against Quidditism'. {\it Australasian Journal of Philosophy}, 78:1, pp.~87-104.\\ 
\\
Br\'ezin, E., Itzykson, C., Parisi, G., and Zuber, J.B.~(1978). ``Planar Diagrams''. {\it Communications in Mathematical Physics}, 59, pp.~35-51.\\
\\
Broad, C.~D.~(1925). {\it The Mind and its Place in Nature}. London: Kegan Paul, Trench, Trubner.\\ 
\\
Butterfield, J. (2011). `Emergence, reduction and supervenience: a varied landscape', \emph{Foundations of Physics}, 41 (6), pp.~920-959. \\
\\
Butterfield, J.~(2011a). `Less is different: emergence and reduction reconciled'. {\it Foundations of Physics}, 41 (6), pp.~1065-1135.\\
\\
Cardy, J.~L.~(Ed.) (1988). {\it Finite-Size Scaling,} vol.~2. Amsterdam: Elsevier Science Publishers.\\
\\
Carnap, R.~(1947). {\em Meaning and Necessity}, Chicago: University of Chicago Press.\\
\\
Castellani, E.~and De Haro, S.~(2019). `Duality, Fundamentality, and Emergence'. Forthcoming in {\it The Foundation of Reality: Fundamentality, Space and Time}, Glick, D., Darby, G., Marmodoro, A.~(Eds.), Oxford University Press. http://philsci-archive.pitt.edu/14494.\\
\\
Chalmers, D.~J.~(2006). `Strong and weak emergence'. {\it The reemergence of emergence}, pp.~244-256.\\
\\
Chang, H.~(2012). {\it Is Water H$_2$O?} Springer Dordrecht Heidelberg New York London.\\
\\
Chandler, D.~(1987). {\it Introduction to Modern Statistical Mechanics}. New York and Oxford: Oxford University Press.\\
\\
Corkum, P.~(2008). `Aristotle on Ontological Dependence'. {\it Phronesis}, 53, pp.~65-92.\\
\\
Crowther, K.~(2016). {\it Effective Spacetime}. Springer International Publishing Switzerland.\\
\\
De Haro, S.~(2015). `Dualities and emergent gravity: Gauge/gravity duality'. {\em Studies in History and Philosophy of Modern Physics}, 59, 2017, pp.~109-125. \\ 
\\
De Haro, S.~(2016). `Spacetime and Physical Equivalence'. Forthcoming in {\it Space and Time after Quantum Gravity}, Huggett, N.~and W\"uthrich, C.~(Eds.). http://philsci-archive.pitt.edu/13243.\\
\\
De Haro, S.~(2019). `Theoretical Equivalence and Duality'. {\it Synthese,} 2019.\\ https://doi.org/10.1007/s11229-019-02394-4.
http://philsci-archive.pitt.edu/16153.\\ 
\\
De Haro, S.~(2019a). `On the Emergence of Masslessness'. In preparation.\\
\\
De Haro, S.~(2019b). `The Emergence of Space, Illustrated by Random Matrix Models'. In preparation.\\
\\
De Haro, S.~and Butterfield, J.~N.~(2017). `A Schema for Duality, Illustrated by Bosonization'. In: {\it Foundations of Mathematics and Physics one century after Hilbert,} pp.~305-376. Kouneiher, J.~(Ed.) Springer, Cham. http://philsci-archive.pitt.edu/13229.\\
\\
De Haro, S.~and Butterfield, J.~N.~(2019). `On Symmetry and Duality'. {\it Synthese,} 2019. https://doi.org/10.1007/s11229-019-02258-x.\\
\\
De Haro, S.~and De Regt, H.~W.~(2018). `Interpreting Theories without Spacetime'. {\it European Journal for Philosophy of Science,} 8 (3), pp.~631-670.  \\
\\
Dieks, D., Dongen, J. van, Haro, S. de~(2015). `Emergence in Holographic Scenarios for Gravity'. 
{\it Studies in History and Philosophy of Modern Physics} 52 (B), pp.~203-216. \\
\\
Dijkgraaf, R., and Vafa, C.~(2002). ``Matrix Models, Topological Strings, and Supersymmetric Gauge Theories''. {\it Nuclear Physics}, B 644, p.~3. 
  [hep-th/0206255].\\
\\
Dizadji-Bahmani, F., Frigg, R., Hartmann, S.~(2010). `Who's afraid of Nagelian reduction?' {\it Erkenntnis}, 73 (3), pp.~393-412.\\
\\
Ellis, G.~(2016). {\it How Can Physics Underlie the Mind?: Top-Down Causation in the Human Context.} Berlin and Heidelberg: Springer-Verlag.\\
\\
Eynard, B., Kimura, T., and Ribault, S.~(2015). ``Random Matrices.'' arXiv preprint arXiv: 1510.04430 [math-ph].\\
\\
Feyerabend, P.~K.~(1963). `How to Be a Good Empiricist---A Plea for Tolerance in Matters Epistemological'. In: Baumrin, B.~(Ed.), {\it Philosophy of Science, The Delaware Seminar,} vol.~2, New York: Interscience Publishers, pp.~3-39. Reprinted in: M.~Curd, and J.~A.~Cover, {\it Philosophy of Science,} New York London: W.~W.~Norton, pp.~922-949.\\
\\
Fine, A.~(1984). `The Natural Ontological Attitude'. In: {\it Scientific Realism}, J.~Leplin (Ed.), pp.~83-107. Berkeley: University of California Press.\\
\\
Fine, K.~(2012). `Guide to Ground'. In: Correia, F., Schnieder, B., {\it Metaphysical Grounding. Understanding the Structure of Reality}. Cambridge University Press.\\
\\
Fletcher, S.~C.~(2016). `Similarity, topology, and physical significance in relativity theory.' {\it The British Journal for the Philosophy of Science}, 67 (2), pp.~365-389.\\
\\
Franklin, A.~and Knox, E.~(2018). `Emergence without limits: The case of phonons'. {\it Studies in History and Philosophy of Modern Physics,} 64, pp.~68-78.\\
\\
Gilmore, R.~(2008). `Lie Groups, Physics, and Geometry'. Cambridge: CUP.\\
\\
Guay, A., Sartenaer, O.~(2016). `A new look at emergence. Or when after is different'. {\it European Journal for Philosophy of Science}, 6 (2), pp.~297-322.\\
\\
Han, D., Kim, Y.~S.~(1981). `Little Group for Photons and Gauge Transformations'. {\it American Journal of Physics}, 49, p.~348.\\
\\
Hempel, C.~(1966). {\it Philosophy of Natural Science}. New York: Prentice-Hall, New York.\\
\\
Hendry, R.F.~(2010). `Ontological reduction and molecular structure'. {\it Studies in History and Philosophy of Modern Physics}, 41, pp.~183-191.\\
\\
Humphreys, P.~(1997). `Properties Emerge'. {\it Philosophy of Science,} 64 (1), pp.~1-17.\\
\\
Humphreys, P.~(2016). {\it Emergence. A Philosophical Account}. Oxford University Press.\\
\\
In\"on\"u, E.~and Wigner, E.~P.~(1953). `On the Contraction of Groups and their Reperesentations'. {\it Proceedings of the National Academy of Sciences},  39 (6), pp.~510-524.\\
\\
Kadanoff, L.~P.~(2013). `Theories of Matter: Infinities and Renormalization'. In: {\it The Oxford Handbook of Philosophy of Physics}, R.~Batterman (Ed.), pp.~141-188. Oxford University Press.\\
\\
Kim, Y.S.~(2001). `Internal Space-time Symmetries of Massive and Massless Particles and their Unification'. {\it Nuclear Physics} B, Proceedings Supplements, 102, pp.~369-376.\\
\\
Kitcher, P.~(1978). `Theories, Theorists and Theoretical Change'. {\it The Philosophical Review}, 87 (4), pp.~519-547.\\
\\
Knox, E.~(2016). `Abstraction and its Limits: Finding Space For Novel Explanation'. {\it No$\hat u$s,} 50 (1), pp.~41-60.\\
\\
Ladyman, J.~and Ross, D.~(2007). {\it Every Thing Must Go. Metaphysics Naturalized.} With Spurrett, D.~and Collier, J.~Oxford: Oxford University Press.\\
\\
Landsman, N.~P.~ (2013). `Spontaneous Symmetry Breaking in Quantum Systems: Emergence or Reduction?' {\it Studies in History and Philosophy of Modern Physics,} 44(4), pp.~379-394.\\
\\
Laughlin, R.~B.~and Pines, D.~(2000). `The Theory of Everything'. {\it Proceedings of the National Academy of Sciences of the United States of America}, 97 (1), pp.~28-31.\\
\\
Lavis, D.~A.~(2017). `The Problem of Equilibrium Processes in Thermodynamics'. {\it Studies in History and Philosophy of Science Part B: Studies in History and Philosophy of Modern Physics,} forthcoming. https://doi.org/10.1016/j.shpsb.2017.07.003.\\
\\
Lewis, D.~K.~(1970). `Index, Context, and Content'. Reprinted in: {\it Papers in Philosophical Logic,} 1998, pp.~21-44. Cambridge: Cambridge University Press.\\
\\
Lewis, D.~K.~(1983). `New Work for a Theory of Universals'. {\it Australasian Journal of Philosophy}, 61 (4), pp.~343-377.\\
\\
Mainwood, P.~(2006). {\it Is More Different? Emergent Properties in Physics}. PhD dissertation, University of Oxford. http://philsci-archive.pitt.edu/8339.\\
\\
Malament, D.~B.~(1986). `Newtonian Gravity, Limits, and the Geometry of Space'. In: {\it From Quarks to Quasars}, R.~G.~Colodny (Ed.), University of Pittsburgh Press.\\
\\
Martin, M.~(1971). `Referential Variance and Scientific Objectivity'. {\it The British Journal for the Philosophy of Science,} 22 (1), pp.~17-26.\\
\\
McLaughlin, B.~P.~(2008). `The Rise and Fall of British Emergentism'. In {\it Emergence: Contemporary Readings in Philosophy and Science}, M.~A.~Bedau and P.~Humphreys (Eds.). Cambridge, MA: The MIT Press.\\
\\
Menon, T.~and Callender, C.~(2013). `Turn and Face the Strange... Ch-Ch-Changes: Philosophical Questions Raised by Phase Transitions'. In: {\it The Oxford Handbook of Philosophy of Physics}, R.~Batterman (Ed.), pp.~189-223. Oxford University Press.\\
\\
Nagel, E.~(1949). `The Meaning of Reduction'. In: {\it Philosophy of Science}, A.~Danto and S.~Morgenbesser (Eds.), pp.~288-312. Meridian Books: Cleveland and New York.\\
\\
Nagel, E.~(1961). {\it The Structure of Science: Problems in the Logic of Scientific Explanation}. New York: Harcourt.\\
\\
Nagel, E.~(1979). `Issues in the Logic of Reductive Explanations'. In: {\it Teleology Revisited,} New York: Columbia University Press, pp.~95-113. Reprinted in: M.~Curd, and J.~A.~Cover, {\it Philosophy of Science,} New York London: W.~W.~Norton, pp.~905-921.\\
\\
Nickles, T.~(1973). `Two Concepts of Intertheoretic Reduction'. {\it The Journal of Philosophy,} 70 (7), pp.~181-201.\\
\\
Norton, J.~D.~(2012). `Approximation and Idealization: Why the Difference Matters', {\it Philosophy of Science}, 79 (2), 207-232.\\
\\
Norton, J.~D.~(2016). `The Impossible Process: Thermodynamic Reversibility'. {\it Studies in History and Philosophy of Science Part B: Studies in History and Philosophy of Modern Physics}, 55, pp.~44-61.\\
\\
O'Connor, T., and Wong, H.~Y.~(2002). `Emergent Properties'. {\it Stanford Encyclopedia of Philosophy}. First published 24 September 2002; substantive revision on 3 June 2015.\\
\\
Onsager, L.~(1944). `Crystal Statistics. I. A Two-Dimensional Model with an Order-Disorder Transition'. {\it Physical Review,} 65 (3)-(4), pp.~117-149.\\
\\
Pitts, J.~B.~(2011). `Permanent Underdetermination from Approximate Empirical Equivalence in Field Theory'. 
{\it The British Journal for the Philosophy of Science}, 62 (2), pp.~259-299.\\
\\
Psillos, S.~(1999). {\it Scientific Realism. How Science Tracks Truth.}  London and New York: Routledge.\\
\\
Quine, W.~V.~O.~(1951). `Ontology and Ideology'. {\it Philosophical Studies. An International Journal for Philosophy in the Analytic Tradition}, 2 (1), pp.~11-15.\\
\\
Quine, W.~V.~O.~(1960). {\it Word and Object}. Cambridge, Massachusetts and London, England: The MIT Press. New edition, 2013.\\
\\
Radder, H.~(2012). `Postscript 2012', in: {\it The Material Realization of Science}, Revised Edition, Springer Dordrecht Heidelberg New York London.\\
\\
Richardson, R.~C.~(2007). `Re-Engineering Philosophy for Limited Beings: Piecewise Approximations to Reality'. {\it Notre Dame Philosophical Reviews}, 2007.12.14.\\
\\
Rueger, A.~(2000). `Physical Emergence, Diachronic and Synchronic'. {\it Synthese}, 124 (3), pp.~297-322.\\
\\
Schaffer, J.~(2009). `On What Grounds What'. In: Chalmers, D.~J., Manley, D., Wasserman, R., {\it Metametaphysics. New Essays on the Foundations of Ontology}. Oxford: OUP.\\
\\
Schaffner, K.~F.~(1967). `Approaches to Reduction'. {\it Philosophy of Science,} 34 (2), pp.~137-147.\\
\\
Schaffner, K.~F.~(2012). `Ernest Nagel and reduction'. {\it The Journal of Philosophy}, 109 (8/9), pp.~534-565.\\
\\
Scheffler, I.~(1967). {\it Science and Subjectivity.} Indianapolis New York Kansas City: The Bobbs-Merrill Company.\\
\\
't Hooft, G.~(1974). ``A Planar Diagram Theory for the Strong Interactions''. {\it Nuclear Physics}, B72, pp.~461-473.\\
\\
van Dongen, J., De Haro, S., Visser, M., Butterfield, J.~(2019). `Emergence and Correspondence for String Theory Black Holes. Forthcoming in {\it Studies in History and Philosophy of Modern Physics.} arXiv:1904.03234 [physics.hist-ph].\\
\\
van Fraassen, B.~C.~(1980). {\it The Scientific Image}. Oxford: Oxford University Press.\\
\\
Wallace, D.~(2010). `Decoherence and Ontology (or: How I learned to stop worrying and love FAPP'. In: {\it Many Worlds?} Saunders, S., Barrett, J., Kent, A., Wallace, D.~(Eds.), pp.~53-72. Oxford: Oxford University Press.\\
\\
Wallace, D.~(2012). {\it The Emergent Multiverse: Quantum Theory According to the Everett Interpretation.} Oxford: Oxford University Press.\\
\\
Wigner, E.~(1939). `On Unitary Representations of the Inhomogeneous Lorentz Group'. Reprinted in: {\it Nuclear Physics B} (Proceedings Supplement), 6, 1989, pp.~9-64.\\
\\
Wimsatt, W.~C.~(1997). `Reductive Heuristics for Finding Emergence'. {\it Philosophy of Science}, 64, pp.~S372-S384.\\
\\
Wimsatt, W.C.~(2007). {\it Re-Engineering Philosophy for Limited Beings. Piecewise Approximations to Reality.} Cambridge, Massachusetts, and London: Harvard University Press.\\
\\
Wong, H.~Y.~(2010). `The Secret Lives of Emergents'. In: A.~Corradini and T.~O'Connor, {\it Emergence in Science and Philosophy,} pp.~7-24. New York and London: Routledge.

\end{document}